\pdfoutput=1

\documentclass[12pt]{article}

\usepackage[ngerman,english,british]{babel}
\usepackage{graphics}
\usepackage{epsfig}
\usepackage{lscape}
\usepackage{amsmath,amsfonts,amssymb}
\usepackage{eurosym}
\usepackage{natbib}
\usepackage{url}
\usepackage{setspace}
\usepackage{caption}
\usepackage{booktabs}
\usepackage{subcaption}
\usepackage{arydshln}
\usepackage{rotating}
\usepackage{pdflscape}
\usepackage{multirow}
\usepackage[hang,flushmargin]{footmisc}

\usepackage{xcolor}

\usepackage[margin=1in]{geometry}

\makeatletter
\def\@sect#1#2#3#4#5#6[#7]#8{\ifnum #2>\c@secnumdepth
     \let\@svsec\@empty\else
     \refstepcounter{#1}\edef\@svsec{\csname the#1\endcsname. \hskip 0.4em}\fi
     \@tempskipa #5\relax
      \ifdim \@tempskipa>\z@
        \begingroup #6\relax
          \@hangfrom{\hskip #3\relax\@svsec}{\interlinepenalty \@M #8\par}%
        \endgroup
       \csname #1mark\endcsname{#7}\addcontentsline
         {toc}{#1}{\ifnum #2>\c@secnumdepth \else
                      \protect\numberline{\csname the#1\endcsname}\fi
                    #7}\else
        \def\@svsechd{#6\hskip #3\relax  %% \relax added 2 May 90
                   \@svsec #8\csname #1mark\endcsname
                      {#7}\addcontentsline
                           {toc}{#1}{\ifnum #2>\c@secnumdepth \else
                             \protect\numberline{\csname the#1\endcsname}\fi
                       #7}}\fi
     \@xsect{#5}}
\makeatother

\makeatletter
\renewcommand{\section}{\@startsection{section}{1}{0mm}{-\baselineskip}{0.25\baselineskip}{\centering\normalfont\normalsize\scshape}}
\renewcommand{\subsection}{\@startsection{subsection}{2}{0mm}{-\baselineskip}{0.25\baselineskip}{\raggedright\normalfont\normalsize\itshape}}
\renewcommand{\subsubsection}{\@startsection{subsubsection}{3}{0mm}{-\baselineskip}{0.25\baselineskip}{\raggedright\normalfont\small\scshape}}
\def\@begintheorem#1#2{\trivlist \item[\hskip \labelsep{\bf #1\ #2:}]\it}
\makeatother

\makeatletter
\def\monthname{\ifcase\month\or
January\or February\or March\or April\or May\or June\or
July\or August\or September\or October\or November\or December\fi}
\makeatother

\renewenvironment{abstract}
 {\begin{center}\normalsize\textsc{Abstract}%
 \end{center}\begin{quote}\normalsize}
 {\end{quote}}

\renewcommand{\appendix}{\footnotesize\parindent 0cm\setlength{\parskip}{\medskipamount}\setcounter{equation}{0}%
\renewcommand{\theequation}{A.\arabic{equation}}}

\newtheorem{proposition}{\small\sc Proposition}
\newtheorem{assumption}{\small\sc Assumption}
\newtheorem{lemma}{\small\sc Lemma}

\newtheorem{definition}{\small\sc Definition}

\usepackage{fancyhdr}
\begin{document}

\begin{titlepage}
\vspace*{0.1cm}

\setcounter{page}{0}

  \begin{center}%
    {\Large

A Generalized Approach to Power Analysis \\ for Local Average Treatment Effects

%Constraints and Enablers for  

 \sc \par}%
       \vskip 3em%
    {\large
     \lineskip .75em%
      \begin{tabular}[t]{c}%
      Kirk Bansak$^*$ \\
     \end{tabular}\par}%
    \vskip 1.5em%
    {\small %First version: June 2016\\
           % This version:
            \monthname \ \number\year \par \vspace{0.1cm}
            Forthcoming, \emph{Statistical Science}}%
      \vskip 1.0em%

%For comments and questions, \\ please e-mail: \texttt{kbansak@stanford.edu}

  \end{center}\par

\vspace{0.25cm}

\begin{abstract}

\noindent 

This study introduces a new approach to power analysis in the context of estimating a local average treatment effect (LATE), where the study subjects exhibit noncompliance with treatment assignment. As a result of distributional complications in the LATE context, compared to the simple ATE context, there is currently no standard method of power analysis for the LATE. Moreover, existing methods and commonly used substitutes---which include instrumental variable (IV), intent-to-treat (ITT), and scaled ATE power analyses---require specifying generally unknown variance terms and/or rely upon strong and unrealistic assumptions, thus providing unreliable guidance on the power of tests of the LATE. This study develops a new approach that uses standardized effect sizes to place bounds on the power for the most commonly used estimator of the LATE, the Wald IV estimator, whereby variance terms and distributional parameters need not be specified nor assumed. Instead, in addition to the effect size, sample size, and error tolerance parameters, the only other parameter that must be specified by the researcher is the compliance rate. Additional conditions can also be introduced to further narrow the bounds on the power calculation. The result is a generalized approach to power analysis in the LATE context that is simple to implement. 

\end{abstract}
\vspace*{0.5cm}
\vspace*{0.35cm}
\vfill
 \footnoterule
    {\footnotesize

\noindent $^*$Assistant Professor, Department of Political Science, University of California, San Diego. \\
\noindent I thank Avidit Acharya, Matthew Blackwell, Justin Grimmer, Jens Hainmueller, Andy Hall, Dominik Hangartner, Kosuke Imai, Guido Imbens, Mike Tomz, Stefan Wager, Teppei Yamamoto, and Xiang Zhou for helpful comments. All errors are my own. \vspace*{0cm}
\noindent\footnotesize }
\end{titlepage}

\newpage
\setcounter{page}{1} \addtolength{\baselineskip}{0.1\baselineskip}

\section{Introduction}

Power analysis has long been recognized as a vital study design tool \citep{cohen1962}. Running simple power analyses provides researchers with concrete and reliable information to help determine their budgetary requirements, choose a sample size, and form reasonable expectations on the magnitude of treatment effects they will be able to detect. This helps researchers avoid an eventuality in which a study has failed to produce meaningful findings not because there is nothing interesting to find but rather due to insufficient power to overcome fundamentally noisy data. 
%, not due to true null effects but simply
The results of a power analysis can also help researchers avoid running certain studies altogether if the costs are simply too prohibitive in light of the probability of successful detection of a meaningful effect. Yet various fields of research are replete with studies that have failed to report power analyses and implemented drastically under-powered designs \citep{tverskykahneman1971, tsangetal2009}. Indeed, many researchers' (and funders') time, energy, and money have been put at risk by neglect of power analysis in the early stages of research design. In practice, however, power analyses can often be challenging to properly implement. 

Consider the standard experimental setting in which units are assigned to one of two conditions, a treatment condition and a control condition, and the goal is to determine whether the treatment has an effect on some outcome variable of interest. Further, it may also be possible that \emph{uptake} of the treatment is not perfectly determined by \emph{assignment} of the treatment, as some units may not comply with their assignment status. To define the causal effects of interest, this study employs the potential outcomes framework presented by \cite{neyman1923applications} and \cite{rubin1974estimating} and postulates a data-generating distribution on quadruples
$(Y_i(0),Y_i(1),D_i(0),D_i(1)) \in \mathbb{R} \times \mathbb{R} \times \{0,1\} \times \{0,1\}$. For any unit $i$, the $Y_i(d)$, $d \in \{0,1\}$, denote the outcome that unit $i$ would exhibit if it undertook treatment status and took the treatment ($d=1$) or if it undertook control status and did not take the treatment ($d=0$). Additionally, the $D_i(z)$, $z \in \{0,1\}$, denote the treatment uptake status that unit $i$ would exhibit if assigned to the treatment condition ($z=1$) or if assigned to the control condition ($z=0$). Throughout this study, we suppose we will observe a sample of $N$ independent and identically distributed units of the form $(Y_i, D_i, Z_i) \in \mathbb{R} \times \{0,1\} \times \{0,1\}$, where for each unit $i$ the $(Y_i(0),Y_i(1),D_i(0),D_i(1))$ is drawn from the distribution noted above, $Z_i$ is a treatment assignment, $D_i = D_i(Z_i)$ is the realized treatment uptake, and $Y_i = Y_i(D_i)$ is the realized outcome. 

In the simple context of full compliance with the treatment assignment (i.e. where all units always take the treatment if assigned to it and do not take the treatment if not assigned to it), then $D_i(0) = 0, D_i(1) = 1$, and hence $D_i = Z_i$, for all $i$. In this setting, researchers are generally interested in the average treatment effect (ATE), which is the difference between the expected outcome units would attain if they took up the treatment and the expected outcome units would attain if they did not take up the treatment. The ATE, denoted here by $\delta$, is defined formally as:
$$\delta = E[Y_i(1) - Y_i(0)]$$
Given full compliance and random assignment of the treatment, consistent and unbiased estimation of $\delta$ is straightforward. Even in this simple case, however, a complication for performing a pre-study power analysis for the test of the null hypothesis that $\delta = 0$ is the need, \emph{a priori}, for estimates of or assumptions about the variances of $Y_i(0)$ and $Y_i(1)$ \citep{bloom2006, dufloetal2007}. 
%While previous studies and/or existing data can often help to inform these variance estimates, there often do not exist any data that are recent or closely related enough to serve as a useful benchmark, particularly when a researcher is interested in a novel outcome variable or a new population of interest. 
This complication is well-known among applied researchers, and fortunately, there also exist fixes for this problem in the full-compliance setting, as will be described later.

Less well-known, however, is the extent to which such complications become exacerbated in the context of estimating treatment effects when the study units exhibit noncompliance with treatment assignment. In this case, even if assigned to the treatment, a unit may not necessarily take the treatment, and vice versa. The distribution of $(Y_i(0),Y_i(1),D_i(0),D_i(1))$ now features four sub-types of units or ``principal strata" that are defined as a function of the $D_i(z)$: ``compliers," defined as the stratum for which $D_i(1)=1$ and $D_i(0)=0$; ``always-takers," for which $D_i(1)=1$ and $D_i(0)=1$; ``never-takers," for which $D_i(1)=0$ and $D_i(0)=0$; and defiers, for which $D_i(1)=0$ and $D_i(0)=1$. In the presence of noncompliance, the ATE generally cannot be identified.\footnote{Noncompliance would not pose a problem for identification of the ATE if units' noncompliance behavior were independent of their potential outcomes. However, in practice this may often be unlikely, as study subjects can be motivated to select into the treatment or control condition based on their expectations of their own potential outcomes under each condition.} However, under a set of assumptions presented by \cite{air1996}, it is possible to identify a ``local average treatment effect" (LATE), which is the ATE for the compliers, or those who take the treatment when assigned to the treatment and do not otherwise.\footnote{The assumptions and identification of the LATE will be discussed more fully in Section III.} The LATE, which will be denoted by $\tau$, is defined formally as:
$$\tau = E[Y_i(1) - Y_i(0) | D_i(1) - D_i(0) = 1]$$

This study considers the problem of performing a power analysis in the presence of noncompliance for the test of the null hypothesis that $\tau = 0$. Due to the existence of multiple principal strata, the possibility of distinct marginal distributional behavior across those strata, the focus on local identification of the average treatment effect for the compliers, and the inability to completely differentiate compliers from other principal strata in observed data, the number of distributional parameters that impact the power vastly proliferates in the LATE context. In fact, the power of the test that $\tau = 0$ not only depends upon the rate of compliance, with lower compliance resulting in lower power, but is also impacted by the different conditional means and variances of the outcome across the principal strata as well as the relative sizes of the principal strata across the distribution (i.e. probabilities of belonging to each stratum). 

As a result, there is currently no standard method of power analysis in the LATE context. In addition, existing methods require specifying distributional assumptions that are difficult to make and/or come with hidden, implicit assumptions about the various principal strata that are unlikely to reflect the reality of one's applied data. Recognizing the complexity of LATE power analysis, some researchers settle for performing scaled ATE or ``intent-to-treat" (ITT) power analyses, discussed later, even when their ultimate estimand of interest is the LATE. This is a precarious practice given that, as will be shown, the results of scaled ATE and ITT power analyses can diverge substantially from the results of a LATE power analysis. This state of affairs is problematic given how common noncompliance is in many research environments, including field experiments, clinical trials, and randomized controlled trials (RCTs) using encouragement designs. These types of studies also tend to be among the most expensive, generating strong incentives for well-calibrated power analyses.

This study introduces a new approach to LATE power analysis employing the Wald IV estimator. Specifically, by using a standardized LATE effect size, this study shows how bounds can be placed on the power of the test of the null hypothesis that $\tau = 0$ whereby neither variance components nor patterns of noncompliance and heterogeneity need to be specified. Instead, in addition to the effect size, sample size, and error tolerance parameters, the only other parameter that must be specified by the researcher is the compliance rate. In contrast, nine other underlying parameters that affect power need not be specified. This study focuses on the Wald IV estimator because it is the most accessible and commonly used estimator of the LATE among applied researchers. In addition, in contrast to other estimators of the LATE, such as those based on maximum likelihood estimation and Bayesian methods \citep[e.g.][]{imbensrubin1997}, the Wald IV estimator is nonparametric.
%and does not require assumptions about the probability distributions underlying the data.

As usual, the effect size and sample size parameters can be isolated in the power analysis introduced in this study, providing ``worst-case-scenario" formulas for minimum detectable effect sizes and required sample sizes. Additional assumptions can also be made to further narrow the bounds on the power calculation to avoid over-conservatism. The result is a generalized approach to power analysis in the LATE context that is simultaneously conservative, disciplined, and simple to implement. As a central reference and summary of the main recommendations, Table \ref{tab:recs} provides the conservative formulas for power, minimum detectable effect size, and required sample size under a variety of scenarios considered in the study, offering researchers a principled strategy for proceeding with conservative power analyses for the LATE. The approach can also be extended to tests in fuzzy regression discontinuity designs \citep{hahnetal2001} that use the instrumental variable (IV) estimator in the discontinuity window around the threshold, as well as quasi-experiments that apply the IV framework to observational data.

To introduce a frame of reference, Section II will briefly discuss power analysis in the standard ATE context with full compliance. Section III will then introduce the LATE, and proceed to highlight problems with existing power analysis methods and general challenges to analyzing power in the LATE context. Sections IV and V will present the new method of LATE power analysis introduced by this study. Section VI will summarize the main results and recommendations, as well as provide an illustration of how the method could be used in practice by applying it in the context of the National JTPA Study. Sections VII and VIII discuss how to extend the framework to allow for covariate adjustment and multi-valued treatments. Section IX concludes.

\section{Power Analysis for Average Treatment Effects}

Consider the goal of understanding how some intervention (a treatment) impacts an outcome of interest in an experimental setting where we can assume full compliance with treatment assignment. As before, suppose we observe a sample of $N$ independent and identically distributed units of the form $(Y_i, D_i, Z_i) \in \mathbb{R} \times \{0,1\} \times \{0,1\}$, where for the $i$th unit $Z_i$ is the treatment assignment, and the outcome $Y_i$ and treatment uptake $D_i$ are generated according to the data-generating distribution of potential outcomes noted in the previous section. Given full compliance, $Z_i = D_i$ for all $i$. Further, let $\widehat{\overline{Y(0)}}$ and $\widehat{\overline{Y(1)}}$ denote the averages of the observed outcomes for the sampled units actually assigned to the control and treatment conditions, respectively, which constitute unbiased and consistent estimates of $E[Y_i(0)]$ and $E[Y_i(1)]$ given random assignment of the treatment and the ``stable unit treatment value assumption" (SUTVA) \citep{rubin1978bayesian, rubin1980randomization, rubin1990comment}. As a result, the difference-in-means estimator, $\hat{\delta} = \widehat{\overline{Y(1)}}-\widehat{\overline{Y(0)}}$, is unbiased and consistent for the average treatment effect (ATE), the true value of which is $\delta = E[Y_i(1) - Y_i(0)]$.

As shown elsewhere \citep{cohen1988, bloom2006, dufloetal2007}, given asymptotic normality of the difference-in-means estimator, power analysis for the test of the null hypothesis that $\delta = 0$ with a two-sided alternative then proceeds with the following equation:
$$\Phi \left(-c^* + \frac{\delta}{\sqrt{V_N}}\right) + \Phi \left(-c^* - \frac{\delta}{\sqrt{V_N}}\right) = 1 - \beta$$
where $\Phi(\bullet)$ denotes the standard normal cumulative distribution function, $\delta$ denotes a hypothesized true ATE value, $V_N$ denotes the sampling variance of the estimator $\hat{\delta}$, $1 - \beta$ denotes the power to correctly reject the null hypothesis ($\beta$ denotes the type-II error rate), and $c^*$ denotes the critical value corresponding to the tolerable type-I error rate ($\alpha$) and hypothesis test type. For the standard two-tailed test of the null hypothesis that $\delta = 0$, $c^* = \Phi^{-1}(1-\frac{\alpha}{2})$. Conventionally, $V_N \equiv \frac{Var(Y_i(1))}{N_1} + \frac{Var(Y_i(0))}{N_0}$, given $N$ units with $N_1$ assigned to treatment and $N_0 = N - N_1$ assigned to control.\footnote{\cite{imbensrubin2015} show that this formula for $V_N$ is conservative given complete randomization of $N$ units with a pre-determined number $N_1$ assigned to treatment and $N_0 = N - N_1$ assigned to control.}

%The standard power formula displayed above can be rearranged depending upon the specific parameter for which the analyst wants to solve. Common parameters to solve for include $\delta$, so that an analyst can recover a minimum detectable effect, and $N$, so that the analyst can recover the necessary sample size. 

%\subsection{ATE Power Analysis Complications}

In order to use the power formula above, the analyst must specify $V_N$, which requires explicitly or implicitly specifying the variances of $Y_i(0)$ and $Y_i(1)$. This requirement presents a possibly serious practical complication. While previous studies and/or existing data can often help to inform these variance specifications, there often do not exist any data that is recent or closely related enough to serve as a useful benchmark, particularly in the case where a researcher is interested in a novel outcome variable or new population of interest. 
%In the case of a binary outcome variable, this problem is partially alleviated by the ability to easily choose a conservative, upper-bound variance. In the case of continuous outcome variables, however, such a mathematical convenience does not exist. 
One might devise an idea, based on theoretical expectations, about what a conservative variance might look like. However, in the very plausible case that these expectations are inaccurate, too high a guess will lead to an overpowered study while too low a guess will lead to an unsuccessful study. In both cases, the researcher's resources are at risk of being wasted.

%Furthermore, lack of reliable knowledge about the potential outcome variances also complicates a researcher's ability to specify meaningful treatment effect values, which is already a subjective task even with known variances. While it is useful to think about effect magnitudes in absolute terms, often times it is difficult to predict what a ``meaningfully large" absolute effect magnitude would be, particularly with less tangible types of outcome variables. Furthermore, from a policy or project evaluation standpoint, it may not be useful to conceptualize the utility or cost-efficiency of an intervention in absolute terms---i.e. without reference to the distribution of the outcome variable of interest. As a result, nonexistent or ill-informed variance estimates doubly hinder proper power analysis by preventing the researcher from specifying reasonable effect magnitudes.

%\subsection{Standardized Effect Sizes}

``Effect sizes" have long been established as the standard solution to this problem in the ATE context with full compliance \citep{cohen1988, bloom2006, dufloetal2007}. In cases where the variances are unknown and/or absolute effect magnitudes are difficult to interpret, a common recommendation is to employ the effect size $\frac{\delta}{\sigma}$---rather than the absolute effect $\delta$---where $\sigma$ is the standard deviation of a reference outcome distribution. In other words, effect sizes are measures of treatment effects that are standardized with reference to the distribution of the outcome variable. Most commonly used is $\sigma = \sqrt{E[Var(Y_i|D_i)]}$, the expected within-group standard deviation of the outcome. By employing effect sizes, the result is that the variance terms in the power formula drop out, thus obviating the inconvenient need to estimate or guess variance values. In addition, there exist general benchmarks for what constitutes a small, medium, and large effect size \citep[e.g.][]{cohen1988}, and meta-analyses within individual fields of study have enabled researchers to develop discipline-specific guidance on effect size significance \citep[Chapter~3]{lipsey1990}.

\section{Power in the LATE Context} \label{s:latepower}

\subsection{Local Average Treatment Effect (LATE)}

In many studies, the subjects exhibit noncompliance: some units assigned to the treatment condition do not take the treatment, and/or some units assigned to the control condition do take the treatment. This problem is pervasive across many research settings---including field experiments, clinical trials, and RCTs using encouragement designs---as subjects often cannot be forced to take the treatment, and some subjects are able to access the treatment even when not assigned to it \citep{gerbergreen2012}. As explained earlier, in the presence of noncompliance, the ATE generally cannot be identified, but
%.\footnote{Noncompliance would not pose a problem for identification of the ATE if units' noncompliance behavior were independent of their potential outcomes. However, in practice this may often be unlikely, as study subjects can be motivated to select into the treatment or control condition based on their expectations of their own potential outcomes under each condition.} However, 
it is possible to identify the local average treatment effect (LATE), which is the ATE for the compliers \citep{air1996}. 
%or those who take the treatment when assigned to the treatment and do not otherwise \citep{air1996}. The compliers are one of three subsets, or principal strata, in the population. The second is the never-takers, or those who never take the treatment regardless of their assignment. The third is the always-takers, or those who always take the treatment regardless of their assignment.\footnote{A fourth stratum, defiers, are those who take the treatment when not assigned to it but do not take the treatment when assigned to it; however, identification of the LATE requires assuming the non-existence of defiers.} 
In the case of one-sided noncompliance, the LATE is the ATE for the treated (given no always-takers) or the ATE for the untreated (given no never-takers).

In their seminal study applying the potential outcomes framework to the identification and estimation of the LATE, \cite{air1996} begin by considering $N$ units indexed by $i$ and defining the potential outcomes $D_i(\mathbf{Z})$ and $Y_i(\mathbf{Z,D})$, where $\mathbf{Z}$ and $\mathbf{D}$ correspond to the $N$-dimensional treatment assignment and uptake vectors across the units. $D_i(\mathbf{Z}) \in \{0,1\}$ denotes the treatment uptake that unit $i$ would exhibit given the full treatment assignment vector, and $Y_i(\mathbf{Z,D}) \in \mathbb{R}$ denotes the outcome that unit $i$ would exhibit given the full treatment assignment and treatment uptake vectors. Note that while this potential outcomes notation differs from that employed earlier in the present study, \cite{air1996} make a set of assumptions that simplify the potential outcomes to the form employed earlier here.
%This differs slightly from the notation introduced earlier, with the treatment assignment and uptake vectors being included here in the definitions of the potential outcomes in order to proceed from the standard identification framework for the LATE introduced by \cite{air1996}. As will become clear, under the assumptions presented in this framework, the potential outcomes simplify to $D_i(Z_i)$ and $Y_i(D_i)$ as in the notation presented earlier. Specifically, 
Specifically, \cite{air1996} introduce the following assumptions:

\begin{assumption}[Stable Unit Treatment Value Assumption (SUTVA)] \label{assump:sutva} 
\hfill \break Let $(\mathbf{Z,D})$ and $(\mathbf{Z',D'})$ be pairs of treatment assignment and uptake vectors. If $Z_i = Z'_i$, then $D_i(\mathbf{Z}) = D_i(\mathbf{Z'})$. If $Z_i = Z'_i$ and $D_i = D'_i$, then $Y_i(\mathbf{Z,D}) = Y_i(\mathbf{Z',D'})$.
%\footnote{The SUTVA assumption is a standard part of the potential outcomes framework, and it implies two things: no interference and only one form of the treatment. Note that the independent and identically distributed sampling employed here implies no interference.}
\end{assumption}
\begin{assumption}[Random Assignment of the Treatment] \label{assump:ra}
\hfill \break $P(\mathbf{Z} = \mathbf{a}) = P(\mathbf{Z} = \mathbf{a}')$ for all $\mathbf{a}$ and $\mathbf{a}'$ such that $\iota^T \mathbf{a} = \iota^T \mathbf{a}'$ where $\iota$ is the N-dimensional column vector with all elements equal to one.
\end{assumption}
\begin{assumption}[Exclusion Restriction] \label{assump:er}
\hfill \break $\mathbf{Y(Z,D)} = \mathbf{Y(Z',D)}$ for all $\mathbf{Z, Z'}$ and for all $\mathbf{D}$.
\end{assumption}
\begin{assumption}[Nonzero Average Causal Effect of Z on D] \label{assump:fse}
\hfill \break $E[D_i(1) - D_i(0)] \neq 0$.
\end{assumption}
\begin{assumption}[Monotonicity] \label{assump:m}
\hfill \break $D_i(1) \geq D_i(0)$ for all i. 
%This implies the non-existence of defiers, or units that take the treatment if not assigned to it and do not take the treatment if assigned to it.
\end{assumption}

As shown by \cite{air1996}, given these assumptions, the potential outcomes for $Y$ are reduced to $Y_i(d)$, $d \in \{0,1\}$, as introduced earlier in this study. The $Y_i(d)$ denote the outcome that unit $i$ would exhibit if unit $i$ assumed treatment status and took the treatment ($d=1$) or if it assumed control status and did not take the treatment ($d=0$), irrespective of all other units. 
%This allows definition of the causal effect of $D$ on $Y$ for unit $i$ as $Y_i(1) - Y_i(0)$. 
Further, the potential outcomes for $D$ are also reduced to the earlier notation, $D_i(z)$, $z \in \{0,1\}$, which denote the treatment uptake status that unit $i$ would exhibit if unit $i$ was assigned to the treatment condition ($z=1$) or assigned to the control condition ($z=0$), irrespective of all other units. 

We can thus postulate, as introduced and defined earlier, a data-generating distribution on the quadruples $(Y_i(0),Y_i(1),D_i(0),D_i(1)) \in \mathbb{R} \times \mathbb{R} \times \{0,1\} \times \{0,1\}$.
Recall that compliers are defined as units for whom $D_i(1) - D_i(0) = 1$. In contrast, always-takers are units for whom $D_i(1) = D_i(0) = 1$, and never-takers are units for whom $D_i(1) = D_i(0) = 0$. Note the existence of defiers, or units for whom $D_i(1) - D_i(0) = -1$ (i.e. units that take the treatment if not assigned to it and do not take the treatment if assigned to it), is ruled out by Assumption \ref{assump:m}.

Following \cite{air1996}, the LATE (denoted here by $\tau$) is defined formally as the ATE, or average causal effect of $D$ on $Y$, for compliers:
$$\tau = E[Y_i(1) - Y_i(0) | D_i(1) - D_i(0) = 1]$$
Under Assumptions \ref{assump:sutva}-\ref{assump:m}, \cite{air1996} show that this estimand is equivalent to the ratio between the average causal effect of $Z$ on $Y$ (intent-to-treat effect, or ITT), which will be denoted by $\gamma$, and the average causal effect of $Z$ on $D$ (first-stage effect), which will be denoted by $\pi$. The first-stage effect is also equivalent to the compliance rate given the assumptions. That is: 
$$\tau = \frac{\gamma}{\pi}$$
$$\gamma = E[Y_i(D_i(1)) - Y_i(D_i(0))]$$
$$\pi = E[D_i(1) - D_i(0)] = P(D_i(1) - D_i(0) = 1)$$

Now suppose we observe a sample of $N$ independent and identically distributed units of the form $(Y_i, D_i, Z_i) \in \mathbb{R} \times \{0,1\} \times \{0,1\}$, where for each unit $i$ the $(Y_i(0),Y_i(1),D_i(0),D_i(1))$ is drawn from the distribution noted above, $D_i = D_i(Z_i)$ given the treatment assignment $Z_i$, and $Y_i = Y_i(D_i)$. Given the assumptions, the LATE can be estimated consistently by the Wald IV estimator, which will be denoted by $\hat{\tau}$:
$$\hat{\tau} = \frac{\widehat{Cov}(Y_i,Z_i)}{\widehat{Cov}(D_i,Z_i)}$$
%where $Y_i$, $D_i$, and $Z_i$ correspond to the observed random variable values for unit $i$. 
where $\widehat{Cov}$ denotes the sample covariance. 

In contrast to other estimators of the LATE, such as those based on maximum likelihood estimation and Bayesian methods \citep[e.g.][]{imbensrubin1997}, the Wald IV estimator is nonparametric and does not require assumptions about the probability distributions underlying the data. The Wald IV estimator is also the most accessible and commonly used estimator of the LATE among applied researchers. The asymptotic variance of the estimator given independent and identically distributed observations, as shown by \cite{imbensangrist1994}, is:
$$V_N^{\hat{\tau}} = \frac{E[\epsilon_i^2 \{ Z_i - E[Z_i] \}^2]}{N Cov^2(D_i,Z_i)}$$
where $\epsilon_i = Y_i - E[Y_i] - \tau(D_i - E[D_i])$.

In the face of noncompliance, researchers often weigh the merits of focusing on the ITT vs. LATE as the ultimate estimand of interest from the perspective of their own research goals and questions \cite[e.g.][]{imbens2014instrumental, kitagawa2014instrumental, swanson2014think, imbens2014rejoinder}. The ITT measures the average effect of treatment \emph{assignment} in the presence of noncompliance. This is an ideal estimand for researchers wishing to understand the overall system-wide effect of introducing an intervention into the study context. However, the ITT does not capture a causal effect of the treatment itself. In contrast, the LATE measures the average causal effect of the treatment \emph{uptake} for the compliers. While the compliers are a subset of the underlying population, note that they are often the sub-population of interest, as they are precisely the subset of individuals who can actually be induced to take (or not take) the treatment. In contrast, it is often not relevant or useful to understand the effect of a treatment for a sub-population who will never end up taking the treatment (or who will always take it no matter what). 

By measuring a causal effect of the treatment, the LATE thereby allows researchers to understand the efficacy of the treatment itself. This can be a critical task for a number of research goals. First, it allows for more direct scientific investigation of the underlying causal phenomenon. Second, it facilitates efforts to improve the design of the intervention such that it becomes more efficacious at the individual level. Third, it is also key for determining the cost-efficiency of the treatment in many contexts. Given that costly interventions often scale proportionately to the number of applications/dosages actually delivered, rather than simply the number assigned, it is crucial to measure the cost-efficiency of delivered treatment applications/dosages, which the LATE allows for but the ITT does not. In short, for studies focused on understanding the efficacy of treatments and measuring their causal effects, the LATE is often a more interesting, informative, and/or policy-relevant estimand.\footnote{Recall, also, that in study designs that ensure the absence of always-takers (never-takers) the LATE becomes the ATE for the treated (untreated).}

\subsection{Proliferation of Parameters Affecting Power for the LATE}

%As alluded to above, the modern appoach to LATE analysis separates the population of interest into three different strata.\footnote{As described earlier, a fourth stratum (defiers) is assumed to not exist.} Conceptualizing these three strata allows for the identification and estimation (via standard IV methods) of a local average treatment effect (LATE), which is the treatment effect for the compliers. The existence of three strata also means there are three distinct sub-populations characterized by different sets of distributional behavior.

In general, we may separate power analysis parameters into three groups: error tolerance parameters, investigation parameters, and distribution parameters. The error tolerance parameters are $\alpha$ (type-I error tolerance) and $\beta$ (type-II error tolerance), where $\beta$ is a parameter only in the case where we are solving for a different parameter rather than calculating the power ($1 - \beta$). The investigation parameters are sample size and effect magnitude/size. These are the parameters of fundamental interest that motivate the use of a pre-study power analysis. Finally, the distribution parameters are the parameters that characterize the distribution(s) of the population(s) of interest. In contrast to the tolerance parameters, which are selected by convention or on a discretionary basis, and the investigation parameters, which the researcher seeks to learn about in order to make research design decisions, the distribution parameters are matters of inconvenience. While they are (usually) not of strict interest to the researcher, the distribution parameters have a dramatic impact on statistical power, and they must be specified at values that are known or believed to reflect reality in order for a power analysis to be properly calibrated and hence informative.

As described earlier, in the standard ATE context with full compliance using absolute effect magnitudes, the power formula requires specification of the variance of the ATE estimator. This variance depends upon two distribution parameters: the potential outcome variances of both the treatment and control conditions. In addition, by employing standardized effect sizes, these distribution parameters can be dispensed with. In the LATE context, a power formula would also entail specifying the variance of the estimator. In contrast to the ATE context, however, this variance depends upon many more distribution parameters in the LATE context. If we consider the compliance rate $\pi$ to be an additional investigation parameter in the LATE context, then there are in fact nine distribution parameters that affect the variance of the Wald IV estimator and hence also affect the power. 

The reason for this proliferation of parameters has to do with marginal distributional heterogeneity across the principal strata.
%, which leads the estimator variance to be affected not only by the conditional outcome variances of the three principal strata, but also the conditional means and the proportion of the population each stratum comprises. 
Specifically, in addition to the investigation parameters, the estimator variance is also affected by: (1) the complier control condition potential outcome mean, $E[Y_i(0) | D_i(1) - D_i(0) = 1]$; (2) the complier control condition potential outcome variance, $Var[Y_i(0) | D_i(1) - D_i(0) = 1]$; (3) the complier treatment condition potential outcome variance, $Var[Y_i(1) | D_i(1) - D_i(0) = 1]$; (4) the never-taker control condition potential outcome mean, $E[Y_i(0) | D_i(1) = D_i(0) = 0]$; (5) the never-taker control condition potential outcome variance, $Var[Y_i(0) | D_i(1) = D_i(0) = 0]$; (6) the always-taker treatment condition potential outcome mean, $E[Y_i(1) | D_i(1) = D_i(0) = 1]$; (7) the always-taker treatment condition potential outcome variance, $Var[Y_i(1) | D_i(1) = D_i(0) = 1]$; (8) the proportion of never-takers, $P(D_i(1) = D_i(0) = 0)$; and (9) the proportion of always-takers, $P(D_i(1) = D_i(0) = 1)$.\footnote{It should be noted that the sum of (8), (9), and $\pi$ must be one, removing a degree of freedom in the specification of distribution parameters. In addition, it should be noted that the treatment condition potential outcome mean for the compliers is not included since it is simply the sum of the compliers' control condition mean and the LATE. Further, treatment (control) condition parameters are not included for the never-takers (always-takers) because the treatment (control) condition never manifests in the data for the never-takers (always-takers) by definition.} These properties are illustrated for the Wald IV estimator in the Supplementary Materials (SM) Appendix B (Tables \ref{tab:sims1}-\ref{tab:sims2}), which presents the results of a series of simulations illustrating the power of the estimator as the marginal distributional characteristics of the principal strata are varied. 

%The simulations (Tables \ref{tab:sims1} and \ref{tab:sims2}) show how, in addition to the investigation parameters, all nine of the distribution parameters affect the power of the estimator. 

%\section{Obstacles to Power Analysis for Local Average Treatment Effects}

\subsection{Limitations of Existing Methods for Power Analysis}

Given the expectation of noncompliance with treatment assignment, a researcher wishing to perform a power analysis in order to inform the study design (e.g. number of subjects) has a few existing options. However, the unique characteristics of the LATE context, namely the existence of multiple principal strata characterized by marginal distributional heterogeneity, significantly limit the reliability of these existing methods.

The first existing option is to apply a standard power analysis to the ITT. This may seem problematic at face value, of course, since the ITT is a different target estimand than the LATE. Indeed, for researchers who intend to focus on and estimate the LATE, the problem with ITT power analyses is that, for a given data-generating distribution, the power to detect non-zero effects for the ITT difference-in-means estimator will deviate from that of estimators of the LATE, as highlighted in previous work by \cite{jo2002}. This phenomenon is illustrated in the SM Appendix B (Table \ref{tab:sims3}), which presents the results of simulations in which the power for the LATE (Wald IV estimator) and ITT (difference-in-means estimator) change at different rates as the simulation specifications are altered. In fact, as the simulations show, the power for tests of the LATE may be higher or lower than the power for tests of the ITT depending upon the heterogeneity across the principal strata. This may be somewhat surprising because the Wald IV estimator is simply a scaled version of the ITT, where the ITT is divided by the compliance rate. However, the compliance rate is a quantity that must also be estimated, and that estimate is generally correlated with the estimate of the ITT, resulting in the Wald IV estimator having distinct statistical properties from the ITT difference-in-means estimator. 

\begin{figure}[ht!]
\begin{center}
\caption{Distribution of Potential Outcomes $Y_i(d)$ in Superpopulation} \label{fig:pstrataexample}
\includegraphics[scale=0.85]{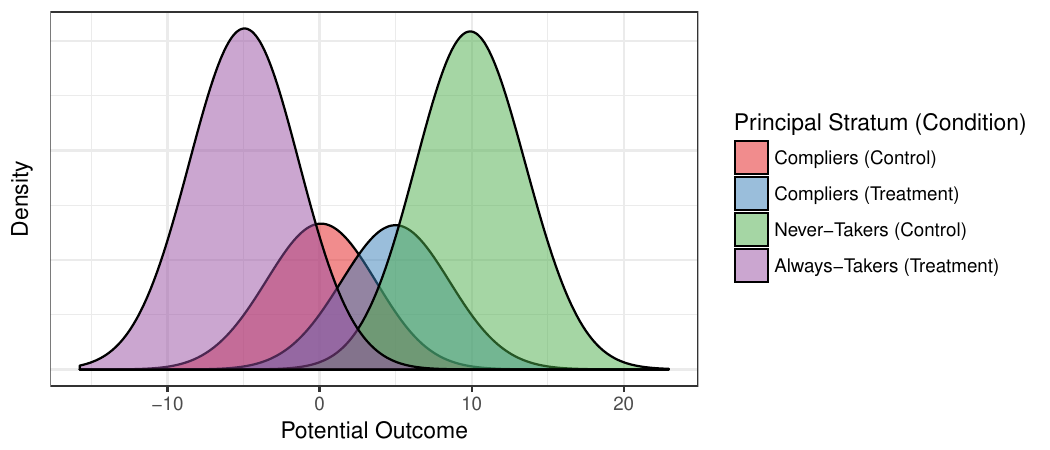}
\end{center}
\end{figure}

To illustrate the problem more vividly, consider a hypothetical superpopulation whose potential outcomes are depicted in Figure \ref{fig:pstrataexample} based on a specific data-generating distribution of the quadruples $(Y_i(0),Y_i(1),D_i(0),D_i(1)) \in \mathbb{R} \times \mathbb{R} \times \{0,1\} \times \{0,1\}$. Compliers comprise 30\% of the superpopulation: $P(D_i(1) - D_i(0) = 1) = 0.3$. Always-takers and never-takers each comprise 35\%: $P(D_i(1) = D_i(0) = 1) = 0.35$ and $P(D_i(1)= D_i(0) = 0) = 0.35$. Finally, defiers do not exist: $P(D_i(1) - D_i(0) = -1) = 0$. Note that compliance rates of 0.3 and lower are prevalent in field experiments, encouragement-based RCTs, and natural experiments.\footnote{For instance, the estimated compliance rate in a vote canvassing field experiment run by \cite{gerbergreen2000} was around 0.3, the estimated compliance rate in an influenza vaccination encouragement design evaluated by \cite{hiranoetal2000} was approximately 0.12, and in a natural experiment evaluated by \cite{angrist1990} on the effect of Vietnam War veteran status on civilian earnings, the estimated compliance rate (effect of draft eligibility on veteran status) ranged from 0.10 to 0.16 for white American citizens born from 1950-1952.} The LATE has a value of 5 in the superpopulation displayed in Figure \ref{fig:pstrataexample}.\footnote{Specifically, the potential outcomes are normally distributed with a variance of 9 and means of 0 and 5 for the compliers under control and under treatment, respectively.} In addition, the never-takers have a slightly higher mean potential outcome value under control and always-takers have a slightly lower mean potential outcome value under treatment, consistent with a common scenario in which subjects with initially high outcome levels have no incentive to take the treatment and subjects with low outcome values are particularly motivated to access the treatment regardless of their assignment.\footnote{Specifically, the potential outcomes for the always-takers (under treatment) and never-takers (under control) are normally distributed with a variance of 9 and means of -5 and 10, respectively.} Note that only the treatment potential outcomes are relevant for the always-takers, and only the control potential outcomes are relevant for the never-takers. 10,000 samples of size 650 were randomly drawn from this superpopulation, and each unit had a probability of 0.5 of being assigned to the treatment. Whether each unit actually took the treatment and its realized outcome were determined jointly by its treatment assignment and the principal stratum to which it belonged, yielding for each sample $650$ independent and identically distributed units of the form $(Y_i, D_i, Z_i) \in \mathbb{R} \times \{0,1\} \times \{0,1\}$. Using the observed values for each sample, tests of the hypotheses that the ITT and LATE are zero were performed with the difference-in-means and Wald IV estimators, respectively, thereby allowing for a simulated comparison of the power of each test.\footnote{The simulated power is the proportion of samples for which the test rejects the null hypothesis of no effect. Two-sided hypothesis tests with $\alpha = 0.05$ were used.} The power for the ITT was approximately 0.77, while the power for the LATE was substantially lower at 0.61. While this is only a single hypothetical data-generating distribution, it conveys an important general message: the results of an ITT power analysis can provide extremely inaccurate guidance (e.g. a miscalibrated sample size recommendation) for researchers planning ultimately to focus on and estimate the LATE. As a general rule, the mismatch between the power for the ITT and power for the LATE will be more pronounced given a lower compliance rate and greater distributional heterogeneity across the principal strata.

A second existing option is a scaled ATE power analysis, which is a commonly used approach in which the results of a standard ATE power analysis are scaled by an appropriate function of the compliance rate. Using this approach, \cite{dufloetal2007} present a formula for computing minimum detectable effects in the presence of noncompliance based on a simple scaling of the standard ATE formula, the result of which follows from the ATE estimator variance being divided by the compliance rate squared. The rationale behind this process is based on the fact that the Wald IV estimator is simply the ratio of the ITT to the compliance rate. However, the scaled ATE power analysis treats the compliance rate as a known value when, as already explained above, the compliance rate must be estimated and that estimate is generally correlated with the ITT estimate. The resulting problem with this approach, which \citeauthor{dufloetal2007} do not make explicit but has been shown elsewhere \citep{baiocchietal2014}, is that a number of strong and unrealistic assumptions are required for this scaling of the standard ATE power analysis to yield the (approximately) correct power for tests using the Wald IV estimator. Specifically, it must be the case that (a) the never-takers have the same mean outcome value as the untreated compliers, (b) the always-takers have the same mean outcome value as the treated compliers, and (c) all groups have the same within-condition outcome variance. If any of those assumptions are violated, the true power of the test of the hypothesis that the LATE equals zero can diverge dramatically from the power implied by this scaled ATE power analysis. SM Appendix B (Table \ref{tab:sims4}) demonstrates this result, illustrating how the scaled ATE power analysis, similar to an ITT power analysis, can provide extremely unreliable guidance on power for the LATE.

Finally, as a third option, power analyses specifically for instrumental-variable (IV) effects have been introduced in the epidemiology literature \citep{pierceetal2011, freemanetal2013, brionetal2013, wang2018sensitivity}. In particular, \cite{freemanetal2013}, \cite{brionetal2013}, and \cite{wang2018sensitivity} all introduce power formulas for IV effects. However, there are two major limitations to the approaches taken by these studies. First, they require specifying a number of variance components, about which a researcher may not have good preexisting knowledge or priors. Second, they proceed from a classic IV perspective and hence neglect the extent to which these variance components depend upon the distinct distributional behavior of the principal strata. 

For instance, the formulas presented by \citeauthor{freemanetal2013} and \citeauthor{brionetal2013} both require specifying $Var(D_i)$.\footnote{$Var(D_i)$ enters both formulas directly as well as through the need to specify $\rho_{DZ}$ (the correlation between $D_i$ and $Z_i$), which can only be mapped from a hypothetical first-stage effect $\pi$ by specifying both $Var(D_i)$ and $Var(Z_i)$.} This presents a challenge given noncompliance with the treatment assignment, as $Var(D_i)$ is a function of both the first stage effect $\pi$ (which is also the compliance rate) and the proportion of always-takers versus never-takers. In other words, to choose an informative value of $Var(D_i)$, one must specify not only a hypothetical compliance rate but also the precise pattern of noncompliance. In addition, \citeauthor{freemanetal2013} also require specifying $Var(Y_i|D_i)$, while \citeauthor{brionetal2013} require specifying the biased asymptotic value of the least squares estimator of the effect of $D$ on $Y$. Finally, the approach taken by \citeauthor{wang2018sensitivity} requires specifying the ITT, standard deviation of the potential outcome under control, standard deviation of the error from regressing the treatment on the instrument, and the correlation between the potential outcome under control and the error from regressing the treatment on the instrument. 
%These are all parameters that a researcher is unlikely to know prior to performing a study, when the power analysis is needed. 
%Finally, both formulas also rely on a simplifying but generally incorrect assumption that $E[\epsilon^2 \{ Z - E[Z] \}^2] = \sigma_2^2 Var(Z)$ in formulating the IV estimator's variance.
%In some research contexts, there may be sufficient preexisting data and previous studies to inform the selection of all the parameters that \citeauthor{freemanetal2013}, \citeauthor{brionetal2013}, and \citeauthor{jiangetal2015} require. Mendelian randomization and other areas of epidemiology may indeed constitute such a case, which explains why \citeauthor{freemanetal2013}, \citeauthor{brionetal2013}, and \citeauthor{jiangetal2015} propose formulas that include these various parameters. 
In many study contexts in the social sciences, medicine, public health, program evaluation, and other fields, the researcher will lack solid estimates or priors on one or more of these parameters. In such contexts, the formulas offered by \citeauthor{freemanetal2013}, \citeauthor{brionetal2013}, and \citeauthor{wang2018sensitivity} cannot be used reliably.

\subsection{General Complications for Designing LATE Power Analyses} \label{ss:complications}

As highlighted above, there are significant limitations and liabilities associated with existing methods of power analysis in the presence of noncompliance. As a result, this is an area in the applied methodological literature that requires new approaches and solutions. Yet there are notable impediments to developing flexible and reliable approaches to power analysis in the LATE context.

%The unique characteristics of the LATE context make designing a power analysis a more complex task than in the ATE context. 

A first complication that has been recognized for some time relates to the local identification of the LATE \citep{jo2002}. As already described, the possibility of heterogeneous potential outcome distributions across the three principal strata (compliers, never-takers, and always-takers) combined with the possibility of different patterns of noncompliance leads to the proliferation of parameters that affect power in the LATE context. Because these parameters jointly factor into the variance of LATE estimators, specifying a hypothetical value for the estimator variance to enable a power analysis involves making explicit or implicit assumptions about all of these parameters.

This first problem leads to a second complication in terms of being able to specify standardized effect sizes in such a way that the variance components of the power analysis drop out. 
%As in the ATE context, the possibility of employing effect sizes suggests a direction for eliminating distributional assumptions. However, 
Whereas in the ATE context given perfect compliance the fix is fairly simple and hence enables the analyst to minimize the number of assumptions that must be made, such a fix has been elusive in the LATE context given imperfect compliance. The result is that existing power analyses in the LATE context, whether analytic or simulation-based, have inconveniently required (explicit or implicit) distributional assumptions that may not match the reality of the data that will eventually be collected.

\section{Introducing a Generalized Approach to Power Analysis for the LATE} \label{s:method}

The remainder of this study introduces a new method of LATE power analysis that addresses the problems described above and provides a more reliable tool than existing methods. The innovation and contribution of this new method is in showing how, by employing effect sizes, bounds can be placed on the power formula whereby neither variance components nor patterns of heterogeneity and noncompliance need to be specified. Instead, in addition to the effect size, sample size, and error tolerance parameters, the only other parameter that must be specified by the researcher is the compliance rate. In other words, only tolerance and investigation parameters must be specified; the analyst need not specify nor even make assumptions about the estimator variance or any of the underlying distribution parameters.

\subsection{Deriving a Modified Power Formula}

As all of the results that follow pertain to estimating the LATE, the standard LATE assumptions (\ref{assump:sutva}-\ref{assump:m}) apply. As before, consider a sample of $N$ independent and identically distributed units of the form $(Y_i, D_i, Z_i) \in \mathbb{R} \times \{0,1\} \times \{0,1\}$, where the LATE will be estimated using the Wald IV estimator, $\hat{\tau} = \frac{\widehat{Cov}(Y_i,Z_i)}{\widehat{Cov}(D_i,Z_i)}$. Similar to the ATE context, the results also invoke the asymptotic normality of the estimator. Hence, the power formula for the test of the null hypothesis that $\tau = 0$ begins as:
$$\Phi \left(-c^* + \frac{\tau}{\sqrt{V_N^{\hat{\tau}}}}\right) + \Phi \left(-c^* - \frac{\tau}{\sqrt{V_N^{\hat{\tau}}}}\right) = 1 - \beta$$
Further assume that assignment to the treatment is randomized with equal probability of being assigned to the treatment and control conditions. (Note that this equal assignment probability assumption will be relaxed later.)

\vspace{.1cm}
\begin{assumption}[Equal Assignment Probability] \label{assump:epa}
\hfill \break $P(Z_i=1) = 0.5$ for all units $i = 1,2,...,N,$
\end{assumption}
\vspace{.1cm}

The LATE power analysis introduced in this study proceeds by defining an effect size of interest. Following conventional practice using effect sizes in ATE power analyses, the effect size is defined in standardized terms with reference to the expected within-assignment-group standard deviation of the outcome:

\vspace{.1cm}
\begin{definition} \label{step4}
Define the effect size of interest as:
$$\kappa = \frac{\tau}{\sqrt{E[Var(Y_i|Z_i)]}}$$
\end{definition}
\vspace{.1cm}

As will be discussed later, defining the effect size in this manner with reference to treatment assignment groups is appealing for a number of reasons. In particular, it provides a structure for determining reasonable effect sizes in advance of a study. In addition, it leads the resulting LATE power analysis derived below to nest the standard ATE power analysis as a special case with full compliance. More discussion is provided in a later section.

By focusing on the effect size, $\kappa$ takes the place of $\tau$ as one of the three investigation parameters, along with $\pi$ and $N$. In order to derive a LATE power formula that does not require specifying distribution parameters, or terms that depend on them, $\frac{\tau}{\sqrt{V_N^{\hat{\tau}}}}$ must be expressed exclusively in terms of investigation parameters. Recall, however, the complexity of the estimator variance: $V_N^{\hat{\tau}} = \frac{E[\epsilon_i^2 \{ Z_i - E[Z_i] \}^2]}{N Cov^2(D_i,Z_i)}$ where $\epsilon_i = Y_i - E[Y_i] - \tau(D_i - E[D_i])$. Further consider that given imperfect compliance in the LATE context, and hence selection into (or out of) the treatment for some subjects, $\epsilon$ is not an intrinsically meaningful disturbance. In particular, $\epsilon$ is not orthogonal to $D$ and hence does not have a conditional expectation of 0; by extension, $E[\epsilon_i^2]$ is not a substantively meaningful term.

As a result of the distributional complexities in the LATE context, it is not possible to derive a point calculation for the power of the Wald IV estimator without specifying its variance or the underlying distribution parameters. However, given Assumption \ref{assump:epa} (equal assignment probability) and Definition \ref{step4}, a set of tight bounds can be derived for the power of the Wald IV estimator. For notational convenience and without loss of generality, assume that $\kappa > 0$ (and hence also $\tau > 0$) is being investigated. Specifically, the following bounds can be put on $\frac{\tau}{\sqrt{V_N^{\hat{\tau}}}}$:

\vspace{.1cm}
\begin{proposition} \label{step6}
Given Assumptions \ref{assump:sutva}-\ref{assump:epa}:
$$\frac{0.5 \kappa \pi \sqrt{N}}{ \sqrt{ 1 + \kappa^2 E[\nu_i^2] +  2 \kappa  \sqrt{E[\nu_i^2]} }}  \leq \frac{\tau}{\sqrt{V_N^{\hat{\tau}}}} \leq \frac{0.5 \kappa \pi \sqrt{N}}{ \sqrt{ 1 + \kappa^2 E[\nu_i^2] -  2 \kappa  \sqrt{E[\nu_i^2]} }}$$
where $\nu_i = D_i - E[D_i] - \pi(Z_i - E[Z_i])$.\footnote{Note that, without loss of generality, it is assumed here that $\kappa > 0$ (and hence also $\tau > 0$) is being investigated. If $\kappa$ and $\tau$ were negative, the inequalities would be reversed.}
\end{proposition}
\vspace{.1cm}

Of particular interest for study design purposes is the lower bound, which can provide the basis for a lower (and hence conservative) bound for the power. Notably, this re-expression leaves only one remaining term that is not an investigation parameter, $E[\nu_i^2]$. However, since $D$ is binary, a practical and conservative re-expression of $E[\nu_i^2]$ can be undertaken by setting $E[\nu_i^2]$ to its largest possible value as a function of $\pi$. The result is a final lower (conservative) bound on $\frac{\tau}{\sqrt{V_N^{\hat{\tau}}}}$:

\vspace{.1cm}
\begin{proposition} \label{step8}
Given Assumptions \ref{assump:sutva}-\ref{assump:epa}:
$$\frac{0.5 \kappa \pi \sqrt{N}}{ 1 + \kappa \sqrt{(0.5 - \frac{\pi}{2})(0.5 + \frac{\pi}{2})} }  \leq \frac{\tau}{\sqrt{V_N^{\hat{\tau}}}}$$
\end{proposition}
\vspace{.1cm}

As can be seen, this final lower bound contains only the three investigation parameters: $\kappa$, $\pi$, and $N$. Furthermore, this lower bound is tight---it cannot be raised without making additional assumptions---thus providing a tight lower bound for the power. In addition, setting $E[\nu_i^2]$ to its largest possible value also results in an approximate (slightly low) upper bound,$ \frac{0.5 \kappa \pi \sqrt{N}}{ \left| 1 - \kappa \sqrt{(0.5 - \frac{\pi}{2})(0.5 + \frac{\pi}{2})} \right| }$, though this quantity is of less practical value for study design than the conservative lower bound.

The bound in Proposition \ref{step8} can be plugged into the power formula $\Phi (-c^* + \tau / \sqrt{V_N^{\hat{\tau}}}) + \Phi (-c^* - \tau / \sqrt{V_N^{\hat{\tau}}})$ to produce a tight lower bound on the power:
$$\Phi \left(-c^* + \frac{0.5 \kappa \pi \sqrt{N}}{ 1 + \kappa \sqrt{(0.5 - \frac{\pi}{2})(0.5 + \frac{\pi}{2})} } \right) + \Phi \left(-c^* - \frac{0.5 \kappa \pi \sqrt{N}}{ 1 + \kappa \sqrt{(0.5 - \frac{\pi}{2})(0.5 + \frac{\pi}{2})} } \right)$$ 
$$\leq 1 - \beta$$
Values of the investigation parameters $\kappa$, $\pi$, and $N$---as well as a type-I error tolerance $\alpha$ to calculate $c^*$---can then be selected in order to calculate the lower bound on the power, $1-\beta$, of the test of the null hypothesis that the LATE is zero. Importantly, there do not exist any variance terms in the modified power formula. As a result, it provides a bound that captures any distributional patterns among all three principal strata. 
%In other words, this modified method of power analysis makes no assumptions, either explicitly or implicitly, about the variance of the estimator or any of the underlying distribution parameters described earlier. 
In other words, the result is a generalized power analysis for the Wald IV estimator of the LATE that is free of additional distributional assumptions and does not require specification of the estimator variance or its underlying distribution parameters.

In addition, as with standard power analyses, the modified power formula can be rearranged to solve for the investigation parameters. Instead of calculating power based on a specific effect size ($\kappa$), compliance rate ($\pi$), and sample size ($N$), it can be more useful to select a desired power level and solve for one of the other parameters by fixing the rest. Of particular interest in this case should be $\kappa$, solving for which will yield the minimum detectable effect size (MDES),\footnote{This term is borrowed from \cite{bloom2006}, who used it in the ATE context as an extension of minimum detectable effects measured in absolute terms \citep{bloom1995}.} and $N$, solving for which will yield the required sample size.

\subsection{Solving for the Minimum Detectable Effect Size}

Again without loss of generality, assume that $\kappa > 0$ is being investigated. Then, for reasonably high levels of power---which includes conventional power levels, such as $0.8$---the second term in the power formula is negligible, simplifying the formula to:
$$\Phi \left(-c^* + \frac{\tau}{\sqrt{V_N^{\hat{\tau}}}}\right) = 1 - \beta$$
Recalling that $c^* = \Phi^{-1}(1-\frac{\alpha}{2})$ for a two-sided test, this can then be re-expressed as:
$$\frac{\tau}{\sqrt{V_N^{\hat{\tau}}}} = \Phi^{-1}\left(1-\frac{\alpha}{2}\right) + \Phi^{-1}(1-\beta)$$
Let $M = \Phi^{-1}(1-\frac{\alpha}{2}) + \Phi^{-1}(1 - \beta)$, which is called the ``multiplier." %As a result, the power formula bounds presented above lead to the following, where again the lower bound is tight and the upper bound is approximate:
%$$\sqrt{ \frac{0.25 \kappa^2 \pi^2 N}{ 1 + \kappa^2 (0.5 - \frac{\pi}{2})(0.5 + \frac{\pi}{2}) +  2 \kappa  \sqrt{(0.5 - \frac{\pi}{2})(0.5 + \frac{\pi}{2})} } }$$ 
%$$\leq M $$ 
%$$\leq \sqrt{ \frac{0.25 \kappa^2 \pi^2 N}{ 1 + \kappa^2 (0.5 - \frac{\pi}{2})(0.5 + \frac{\pi}{2}) -  2 \kappa  \sqrt{(0.5 - \frac{\pi}{2})(0.5 + \frac{\pi}{2})} } }$$
$M$ can then be plugged in for $\frac{\tau}{\sqrt{V_N^{\hat{\tau}}}}$ in the bounds presented above. This allows $\kappa$ to then be isolated such that the MDES can be computed as a function of the other parameters. This results in a tight upper bound on the MDES, corresponding to the tight lower bound on the power, that can then be used as a conservative value for study design purposes:
$$\kappa \leq \frac{2M}{\pi \sqrt{N} - 2M \sqrt{(0.5 - \frac{\pi}{2})(0.5 + \frac{\pi}{2})}}$$
As before, an approximate lower bound on the MDES can also be expressed in the same manner, though the MDES lower bound is of less practical value than the MDES upper bound for study design.\footnote{The MDES approximate lower bound is $\frac{2M}{\pi \sqrt{N} + 2M \sqrt{(0.5 - \frac{\pi}{2})(0.5 + \frac{\pi}{2})}} \leq \kappa$.}

%$$\pmb{\kappa_{High}} = \frac{2M}{\pi \sqrt{N} - 2M \sqrt{(0.5 - \frac{\pi}{2})(0.5 + \frac{\pi}{2})}}$$
%$$\kappa_{Low} = \frac{2M}{\pi \sqrt{N} + 2M \sqrt{(0.5 - \frac{\pi}{2})(0.5 + \frac{\pi}{2})}}$$
%$\kappa_{High}$ represents the tight conservative bound corresponding to the tight lower bound on the power.

\subsection{Solving for the Sample Size}
Instead of isolating $\kappa$ as above, $N$ can be isolated in order to solve for the required sample size. Continuing to assume without loss of generality that $\kappa > 0$ is being investigated, the following is the tight upper bound on the required sample size corresponding to the tight lower bound on the power:
$$N \leq \frac{4 M^2 \left( 1 + \kappa \sqrt{(0.5 - \frac{\pi}{2})(0.5 + \frac{\pi}{2})} \right)^2 }{\kappa^2 \pi^2}$$
This upper bound can then be used as a conservative value for study design purposes. Once again, an approximate lower bound on the required sample size can be similarly expressed, with the same caveat that such a quantity holds less practical value than the upper bound.\footnote{The required sample size approximate lower bound is $\frac{4 M^2 \left( 1 - \kappa \sqrt{(0.5 - \frac{\pi}{2})(0.5 + \frac{\pi}{2})} \right)^2 }{\kappa^2 \pi^2} \leq N$.}

%$$\pmb{N_{High}} = \frac{4 M^2 \left( 1 + \kappa \sqrt{(0.5 - \frac{\pi}{2})(0.5 + \frac{\pi}{2})} \right)^2 }{\kappa^2 \pi^2}$$
%$$N_{Low} = \frac{4 M^2 \left( 1 - \kappa \sqrt{(0.5 - \frac{\pi}{2})(0.5 + \frac{\pi}{2})} \right)^2 }{\kappa^2 \pi^2}$$
%$N_{High}$ represents the tight conservative bound corresponding to the tight lower bound on the power.

%$$N_{Low} = \frac{2 \kappa^3 M^2 \pi^2 \sqrt{-(\pi-1)(\pi+1)} + \kappa^2 M^2 \sqrt{4-\kappa^2 \pi^2} - 2 \kappa^2 M^2 \pi^2 \sqrt{4 - \kappa^2 \pi^2} + 4 M^2 \sqrt{4-\kappa^2 \pi^2} - 8 \kappa M^2 \sqrt{-(\pi-1)(\pi+1)}}{\kappa^2 \pi^2 \sqrt{4 - \kappa^2 \pi^2}}$$
%$$N_{High} = \frac{-2 \kappa^3 M^2 \pi^2 \sqrt{-(\pi-1)(\pi+1)} + \kappa^2 M^2 \sqrt{4-\kappa^2 \pi^2} - 2 \kappa^2 M^2 \pi^2 \sqrt{4 - \kappa^2 \pi^2} + 4 M^2 \sqrt{4-\kappa^2 \pi^2} + 8 \kappa M^2 \sqrt{-(\pi-1)(\pi+1)}}{\kappa^2 \pi^2 \sqrt{4 - \kappa^2 \pi^2}}$$

\subsection{Narrowing the Bounds} \label{ss:narrowbounds}

By providing a strict lower bound on the power, the method presented above offers a disciplined and reliable means of performing a conservative power analysis for the LATE. However, there is a tradeoff between conservatism and efficiency. If the lower bound is too conservative, it will lead to underestimation of the power and hence overestimation of the MDES and sample size required. This could then result in sub-optimal outcomes, such as a study being over-funded to achieve the conservative sample size or perhaps not funded at all if the sample size requirements exceed the financial resources available. As a result, it would be useful to narrow the bounds on the power formula where possible.

Continuing to assume without loss of generality that $\kappa > 0$, it can be shown that the lower bound on the power formula can be substantially raised when $\bar{Y}_{NT} \leq \bar{Y}_{C} \leq \bar{Y}_{AT}$, where $\bar{Y}_{C}$, $\bar{Y}_{NT}$, and $\bar{Y}_{AT}$ denote the expected realized outcome value for compliers, never-takers, and always-takers (e.g. $\bar{Y}_{C} = E[Y_i|Complier] = E[Y_i|D_i(1) - D_i(0) = 1]$).

\vspace{.1cm}
\begin{assumption}[Ordered Means] \label{assump:om}
\hfill \break $\bar{Y}_{NT} \leq \bar{Y}_{C} \leq \bar{Y}_{AT}$ where $\bar{Y}_{C}$, $\bar{Y}_{NT}$, and $\bar{Y}_{AT}$ denote the expected realized outcome value for compliers, never-takers, and always-takers.
\end{assumption}
\vspace{.1cm}

In the case of one-sided noncompliance, Assumption \ref{assump:om} (ordered means) can be simplified to $\bar{Y}_{NT} \leq \bar{Y}_{C}$ or $\bar{Y}_{C} \leq \bar{Y}_{AT}$, depending upon the direction of noncompliance. It should also be noted that, in the case where noncompliance is almost one-sided (i.e. very few always-takers or very few never-takers), the sparse principal stratum will have only a negligible impact on estimation. Thus, as a practical matter, Assumption \ref{assump:om} can be simplified to $\bar{Y}_{NT} \leq \bar{Y}_{C}$ or $\bar{Y}_{C} \leq \bar{Y}_{AT}$ as long as the sparse principal stratum is deemed sufficiently small.

As a result, if the researcher is comfortable making Assumption \ref{assump:om}, then the lower bound of the power formula can be raised by using the following:

\vspace{.1cm}
\begin{proposition} \label{newbound}
Given Assumptions \ref{assump:sutva}-\ref{assump:om}:
$$ \frac{0.5 \kappa \pi \sqrt{N}}{\sqrt{ 1 + \kappa^2 (0.5 - \frac{\pi}{2})(0.5 + \frac{\pi}{2}) }}  \leq \frac{\tau}{\sqrt{V_N^{\hat{\tau}}}}$$
\end{proposition}
\vspace{.1cm}
Plugging this into the power formula yields:
$$\Phi \left(-c^* + \frac{0.5 \kappa \pi \sqrt{N}}{\sqrt{ 1 + \kappa^2 (0.5 - \frac{\pi}{2})(0.5 + \frac{\pi}{2}) }} \right) + \Phi \left(-c^* - \frac{0.5 \kappa \pi \sqrt{N}}{\sqrt{ 1 + \kappa^2 (0.5 - \frac{\pi}{2})(0.5 + \frac{\pi}{2}) }} \right) \leq 1 - \beta$$
By the same process described earlier, it is possible to solve for $\kappa$ and $N$ to derive new (lowered) upper bounds on the MDES and required sample size:
$$\kappa^* \leq \frac{2 M}{\sqrt{N \pi^2 - 4 M^2 (0.5 - \frac{\pi}{2})(0.5 + \frac{\pi}{2})}}$$
$$N^* \leq  \frac{4 M^2 ( 1 + \kappa^2 (0.5 - \frac{\pi}{2})(0.5 + \frac{\pi}{2}))}{\kappa^2 \pi^2}$$
where again $M = \Phi^{-1}(1-\frac{\alpha}{2}) + \Phi^{-1}(1 - \beta)$.

When would Assumption \ref{assump:om} (ordered means) be reasonable? Roughly speaking, there are two factors to consider when assessing the plausibility of this assumption. The first relates to effect heterogeneity. Specifically, it should be the case that always-takers (never-takers) select into (out of) the treatment because treatment uptake for them is associated with effects that are larger (smaller) than the average treatment effect for the compliers, or at least similarly sized. For instance, in the case of a positive and beneficial treatment, we must expect the noncomplying study subjects to be sufficiently rational that they are selecting into (out of) the treatment in anticipation of a particularly good (bad) effect on their outcome. Alternatively, selection into and out of the treatment could also be made for arbitrary reasons that are uncorrelated with individual effects. The second factor relates to baseline outcome levels in the absence of the treatment. Specifically, we must expect that always-takers (never-takers) do not have baseline outcome levels that are particularly low (high) compared to that of the compliers.

%Given that this derivation is based, WLOG, on a positive treatment effect, it may be reasonable to expect that treatment uptake has a positive relationship with the outcome regardless of whether the uptake is the result of assignment (compliance) or not (noncompliance). Further, since there is full treatment uptake among the always-takers, partial uptake among the compliers, and no uptake among the never-takers, this would suggest that the inequality presented above is a reasonable condition. Of course, such an assumption should not be taken lightly or in the absence of theory.

Precisely when the assumption should be expected to hold, in light of the two factors described above, will certainly be context dependent. However, specific research design steps can be taken to increase its plausibility. First, studies can often be designed so as to exclude one type of noncompliance. Indeed, many experiments are designed to prevent those not assigned to the treatment from accessing it and thus ensuring the absence of always-takers. By achieving one-sided noncompliance, the analyst need only consider two principal strata, rather than three. The well-known National JTPA Study is one such example \citep{bloometal1997}. In this experimental study, subjects were randomly assigned such that they were either given an offer to enroll in a job training program (assigned to treatment) or excluded from participating in the training for an 18-month period (assigned to control). However, many subjects given access to the job training program decided not to receive the training, resulting in a large chunk of never-takers. While there were some enterprising individuals who gained access to the job training in spite of not being assigned to it, their numbers were so small that there was virtually one-sided noncompliance. 

Another research design step that can be taken is to impose reasonable restrictions on the study population of interest to ensure more similar baseline outcome levels across the principal strata. Again, the JTPA experiment is illustrative, as eligibility to participate in the experiment was restricted to those with economic disadvantages and barriers to employment.  Had such restrictions not been made, the study may have included employed and/or higher income professionals who likely would have opted out of the job training program regardless of treatment assignment, boosting the baseline economic outcome levels of the never-takers. 

In fact, as shown in the SM Appendix C, the final results from the JTPA experiment were consistent with the ordered means assumption in terms of an outcome variable that measured the participants' earnings in the 30-month period following their random assignment. Appendix C also presents the results of two other studies that were consistent with the ordered means assumption. One is a vote-canvassing field experiment \citep{gerbergreen2000}. The other is a fuzzy regression discontinuity design on the effect of naturalization on political integration \citep{hainhanpiet2015}. That the ordered means assumption was met in all three of these studies, which involved distinct study designs and research topics, demonstrates the plausibility of this assumption in various research domains.

In contrast, however, another common scenario in field experiments and encouragement-based RCTs is where subjects with initially high outcome levels have no incentive to take the treatment and subjects with low outcome values are particularly motivated to access the treatment regardless of their assignment. This pattern can lead to an ordering of the principal strata means that is the opposite of the ordered means assumption, as illustrated earlier in Figure \ref{fig:pstrataexample}. If the researcher believes such a scenario to be possible, the ordered means assumption should not be applied, and the conservative lower bound on the power as reflected by Proposition \ref{step8} should be used.

% almost all of the guesswork involved in power analysis. Whereas in a basic LATE power analysis framework, one would have to specify various unknowns: one-sided or two-sided noncompliance?, variance(s) of Y, and compliance rate. In my framework, the only thing the analyst needs to fix is the compliance rate.

\subsection{Discussion on Effect Sizes}

As explained earlier, the effect size of interest in this study ($\kappa$) is defined similarly to the way effect sizes are conventionally defined in ATE power analyses, with reference to the expected standard deviation of the outcome within treatment assignment groups. The difference, of course, is that full compliance is assumed in the ATE case, and hence treatment assignment is equivalent to treatment uptake. Nonetheless, the treatment assignment groups remain conceptually and practically useful reference groups in the LATE case for several reasons.

First, the treatment assignment groups are a mathematically natural reference group, allowing standard ATE power analysis results to nest as a special case within the LATE power formula presented in this study. Consider that in the special case of full compliance ($\pi=1$), the LATE becomes the ATE. Further, given $\pi=1$, the LATE power bounds presented in this study, as laid out in Proposition \ref{step8}, are simplified to a single value:
$$1 - \beta = \Phi \left(-c^* + \frac{\kappa \sqrt{N}}{2} \right) + \Phi \left(-c^* - \frac{\kappa \sqrt{N}}{2}\right)$$
Solving for the minimum detectable effect size and required sample size yields $\kappa = \frac{2M}{\sqrt{N}}$ and $N = \frac{4M^2}{\kappa^2}$, where again $M = \Phi^{-1}(1-\frac{\alpha}{2}) + \Phi^{-1}(1 - \beta)$. These results are identical to the conventional ATE power analysis results given asymptotic normality of the estimator and equal probability of assignment to treatment and control \citep{cohen1988, bloom2006}.

Second, the treatment assignment groups contain a natural reference point for defining a standardized effect size. In particular, the distribution of the outcome under assignment to control represents a natural state of the world in the absence of intervention, and hence $Var(Y_i|Z_i=0)$ is one of the few baseline values that can often be reliably measured or estimated in advance of a study by analyzing data on the baseline population.\footnote{In contrast, noncompliance leads to non-randomization of $D$, which means $Var(Y_i|D_i=0)$ will not be accurately reflected by pre-study estimates.} Further, while $Var(Y_i|Z_i=1)$ cannot be measured in advance, it may be reasonable to assume it is relatively close in value to $Var(Y_i|Z_i=0)$. In such cases, $E[Var(Y_i|Z_i)] \approx Var(Y_i|Z_i=0)$, and hence the effect size of interest is defined (approximately) with reference to a naturally occurring distribution that is measurable prior to study implementation.

Third, because $Var(Y_i|Z_i=0)$ may be measurable or estimable in advance, this allows for approximate mapping of effect sizes to absolute effects in the power formula. As long as the researcher is comfortable assuming that $Var(Y_i|Z_i=1)$ will not diverge substantially from $Var(Y_i|Z_i=0)$, then the researcher may estimate $\hat{\omega}_0 = \widehat{\sqrt{Var(Y_i|Z_i=0)}}$, use that estimate as an approximate value for $\sqrt{E[Var(Y_i|Z_i)]}$, and hence replace $\kappa$ with $\frac{\tau}{\hat{\omega}_0}$. The results presented above could then be modified to solve for a minimum detectable absolute effect (i.e. solve for $\tau$ itself) or solve for the required sample size in terms of $\tau$.

Irrespective of the availability of reliable estimates for $\hat{\omega}_0$, researchers may also determine a target MDES by surveying previous studies and meta-analyses within their own fields of study \citep[e.g.][Chapter~3]{lipsey1990}. In his seminal presentation of the topic, \cite{cohen1988} offered the conventional benchmarks in the social and behavioral sciences of 0.2, 0.5, and 0.8 as small, medium, and large effect sizes, respectively. These general conventions may be useful as rough guidance. However, what is considered a small or large effect size inevitably varies across disciplines and research topics. Accordingly, it is advisable for the researcher to more carefully characterize the effect size scale within the research context at hand, in consultation with relevant data from previous studies and meta-analyses, as is the case for any power analysis irrespective of study design and compliance levels.

\subsection{Comparing the Bounds to Simulations}

\begin{figure}[ht!]
\begin{center}
\caption{Simulated Power vs. Analytic Bounds} \label{fig:simsbounds}
\includegraphics[scale=0.8]{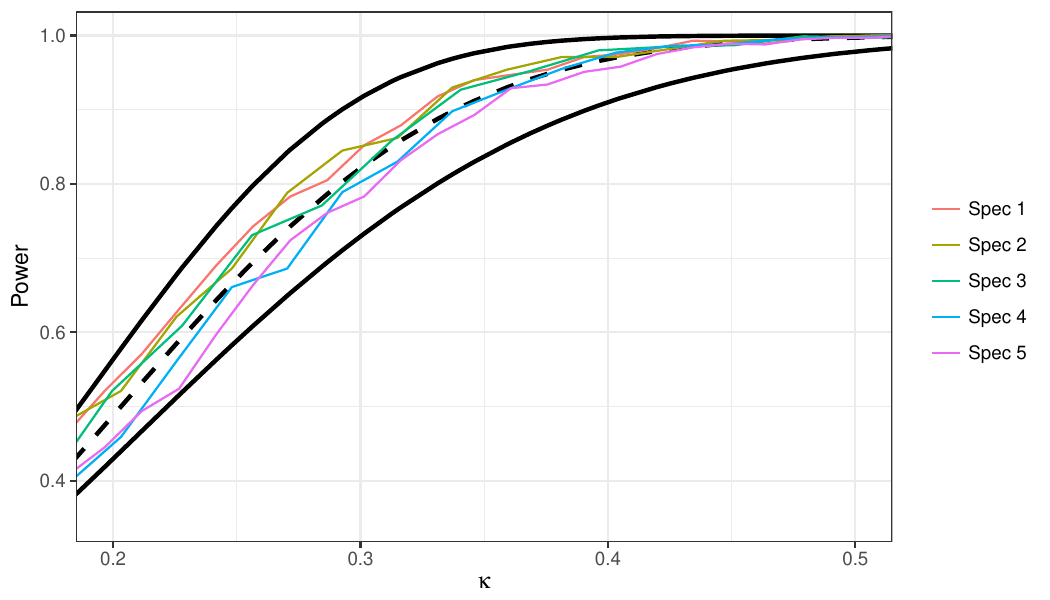}
\end{center}
\end{figure}

To further validate the LATE power bounds derived in this study, Figure \ref{fig:simsbounds} compares the bounds to simulated power curves, where power is plotted as a function of $\kappa$. As in the simulation presented earlier, the simulations presented here also each specify a data-generating distribution of the quadruples $(Y_i(0),Y_i(1),D_i(0),D_i(1)) \in \mathbb{R} \times \mathbb{R} \times \{0,1\} \times \{0,1\}$, randomly draw from that distribution, and randomize the treatment assignment variable, generating samples of independent and identically distributed units of the form $(Y_i, D_i, Z_i) \in \mathbb{R} \times \{0,1\} \times \{0,1\}$.

The solid black lines denote the analytic upper and lower bounds of the power, while the dashed black line denotes the alternative lower bound under Assumption \ref{assump:om} (ordered means). The colored lines denote the power curves that were simulated by specifying the full set of investigation and distribution parameters. For all of the curves (analytic and simulated), the following parameters are fixed: $\pi = 0.5$, $N=1500$, $\alpha = 0.05$. In addition, for the simulated power curves, the following seven of the nine distribution parameters are fixed: $E[Y_i(0)|Complier] = 0$, $Var(Y_i(0)|Complier) = 64$, $Var(Y_i(1)|Complier) = 64$, $Var(Y_i(0)|NeverTaker) = 144$, $Var(Y_i(1)|AlwaysTaker) = 16$, $P(NeverTaker) = 0.25$, and $P(AlwaysTaker) = 0.25$. In contrast, the final two distribution parameters, $E[Y_i(0)|NeverTaker]$ and $E[Y_i(1)|AlwaysTaker]$, vary across the five different simulation specifications shown in different colors. The five sets of values of $E[Y_i(0)|NeverTaker]$ and $E[Y_i(1)|AlwaysTaker]$, starting with the first specification, are as follows: $(-20,20)$, $(-10,10)$, $(-3,3)$, $(10,-10)$, and $(20,-20)$.\footnote{In these simulations, the underlying super populations were generated with normally distributed potential outcomes with means and variances according to the specifications described here.} This ensures that the simulation includes specifications that both do and do not meet Assumption \ref{assump:om} (ordered means), and hence allows for detailed evaluation of the bounds.

Figure \ref{fig:simsbounds} provides a simple demonstration of the performance of the power bounds presented in this study. As can be seen, the simulated curves fall within the analytic bounds denoted by the solid black lines.
%\footnote{It appears, however, that the simulated power curves fall slightly outside the bounds at low levels of power. This is not necessarily surprising given that the power formula relies upon asymptotic normality, the estimator is likely to exhibit somewhat irregular tail behavior in finite samples, and the estimator's variance must itself be estimated.
%whereas the simulations are based upon finite samples and hence deal with random variables not exactly distributed standard normal (the tails are probably fatter).
%} 
Furthermore, the alternative lower bound (the dashed black line) also bounds the appropriate simulated curves. For specifications 1 and 2, Assumption \ref{assump:om} (ordered means) is met by design at all values of $\kappa$, and hence the alternative lower bound applies. Accordingly, the curves for these specifications lie above the alternative lower bound. In contrast, the ordered means assumption is violated by specifications 4 and 5. Thus, it is no surprise that the curves for these specifications lie below the alternative lower bound. Additional graphical illustrations of the relationships between power and the investigation parameters are provided in the SM Appendix D.

\section{Relaxing the Equal Assignment Probability Assumption}

In some situations, the researcher may have reason put an unequal probability on assignment to the treatment and control conditions. For instance, treatment assignment/encouragement may be costly. For such cases, it will be useful to relax Assumption \ref{assump:epa}.

\subsection{Results with $P(Z_i=1) = p_z$}

In the IV-LATE literature, the simplifying assumption of homoskedasticity that $E[\epsilon_i^2 | Z_i] = E[\epsilon_i^2]$ is often made. 
%In other words, the error variance in equation (\ref{eq2}) does not depend upon $Z$. 
While there may be few cases in which this assumption is likely to hold exactly, it is often sufficiently reasonable such that it does not substantially affect statistical inference. The assumption that $E[\epsilon_i^2 | Z] = E[\epsilon_i^2]$ can be useful here.

\vspace{.1cm}
\begin{assumption}[Homoskedasticity] \label{assump:homosked}
\hfill \break $E[\epsilon_i^2 | Z_i] = E[\epsilon_i^2]$
\end{assumption}
\vspace{.1cm}
However, because the simplifying assumption that $E[\epsilon_i^2 | Z_i] = E[\epsilon_i^2]$ is not a conservative one, it is useful to induce conservatism elsewhere. In order to do this, we can consider the limiting value of $E[\nu_i^2]$ at $0.25$.

Continuing to assume without loss of generality that $\kappa > 0$ (and hence also $\tau > 0$) is under investigation, the following lower bound on $\frac{\tau}{\sqrt{V_N^{\hat{\tau}}}}$ then follows without making Assumption \ref{assump:epa} (equal assignment probability):

\vspace{.1cm}
\begin{proposition} \label{relaxedbounds}
Given Assumptions \ref{assump:sutva}-\ref{assump:m} and \ref{assump:homosked}, 
%$E[\nu_i^2] = 0.25$, 
and any value $p_z = P(Z_i=1)$:
$$\frac{\kappa \pi \sqrt{p_z (1-p_z) N}}{1 + 0.5 \kappa} \leq \frac{\tau}{\sqrt{V_N^{\hat{\tau}}}}$$
%$$\sqrt{ \frac{p_z (1-p_z) \kappa^2 \pi^2 N}{ 1 + 0.25 \kappa^2 +  \kappa }} \leq \frac{\tau}{\sqrt{V_N^{\hat{\tau}}}}$$
%$$\sqrt{ \frac{p_z (1-p_z) \kappa^2 \pi^2 N}{ 1 + 0.25 \kappa^2 +  \kappa }} \leq \frac{\tau}{\sqrt{V_N^{\hat{\tau}}}} \leq \sqrt{ \frac{p_z (1-p_z) \kappa^2 \pi^2 N}{ 1 + 0.25 \kappa^2 -  \kappa }}$$
\end{proposition}
\vspace{.1cm}

As before, to derive the lower bound on the power, the bound above can simply be plugged into the power formula $\Phi (-c^* + \tau / \sqrt{V_N^{\hat{\tau}}}) + \Phi (-c^* - \tau / \sqrt{V_N^{\hat{\tau}}})$. Again, $\kappa$ and $N$ can be isolated such that the MDES and required sample size can be computed as a function of the other parameters. The resulting conservative upper bounds on the MDES and required sample size are as follows:
$$\kappa \leq \frac{2M}{2 \pi \sqrt{N p_z (1-p_z)} - M}$$
%$$\pmb{\kappa_{High}} = \frac{2M}{2 \pi \sqrt{N p_z (1-p_z)} - M}$$
%$$\kappa_{Low} = \frac{2M}{2 \pi \sqrt{N p_z (1-p_z)} + M}$$
%, and $\kappa_{High}$ represents the conservative bound.
%$N$ can also be isolated to solve for the required sample size. The resulting bounds on the sample size are as follows:

$$N \leq \frac{M^2 (1 + 0.5 \kappa)^2}{p_z (1-p_z) \kappa^2 \pi^2}$$
%$$\pmb{N_{High}} = \frac{M^2 (1 + 0.25 \kappa^2 + \kappa)}{p_z (1-p_z) \kappa^2 \pi^2}$$
%$$\pmb{N_{High}} = \frac{M^2 (1 + 0.5 \kappa)^2}{p_z (1-p_z) \kappa^2 \pi^2}$$
%$$N_{Low} = \frac{M^2 (1 + 0.25 \kappa^2 - \kappa)}{p_z (1-p_z) \kappa^2 \pi^2}$$
%$$N_{Low} = \frac{M^2 (1 - 0.5 \kappa)^2}{p_z (1-p_z) \kappa^2 \pi^2}$$
%$N_{High}$ represents the conservative bound.
where again $M = \Phi^{-1}\left(1-\frac{\alpha}{2}\right) + \Phi^{-1}\left(1 - \beta \right)$.

\subsection{Narrowing the Bounds while Relaxing the Equal Assignment Probability Assumption}

As before, the lower bound of the power can be increased under Assumption \ref{assump:om} (ordered means). The result is the following alternative lower bound.

\vspace{.1cm}
\begin{proposition} \label{newnewbound}
Given Assumptions \ref{assump:sutva}-\ref{assump:m} and \ref{assump:om}-\ref{assump:homosked}, 
%$E[\nu_i^2] = 0.25$, 
and any value $p_z = P(Z_i=1)$:
$$\frac{\kappa \pi \sqrt{p_z (1-p_z)N}}{\sqrt{1 + 0.25 \kappa^2}} \leq \frac{\tau}{\sqrt{V_N^{\hat{\tau}}}}$$
%$$\sqrt{ \frac{p_z (1-p_z) \kappa^2 \pi^2 N}{ 1 + 0.25 \kappa^2 }} \leq \frac{\tau}{\sqrt{V_N^{\hat{\tau}}}}$$
\end{proposition}
\vspace{.1cm}

%Again, this bound above can be plugged into the power formula $\Phi (-c^* + \tau / \sqrt{V_N^{\hat{\tau}}}) + \Phi (-c^* - \tau / \sqrt{V_N^{\hat{\tau}}})$. 
Solving for $\kappa$ and $N$ to derive alternative (lowered) MDES and required sample size upper bounds leads to the following:
$$\kappa^* \leq \frac{2M}{\sqrt{4 \pi^2 N p_z (1-p_z) - M^2}}$$
$$N^* \leq \frac{M^2 (1 + 0.25 \kappa^2)}{p_z (1-p_z) \kappa^2 \pi^2}$$
where again $M = \Phi^{-1}(1-\frac{\alpha}{2}) + \Phi^{-1}(1 - \beta)$

Appendix E in the SM presents results comparing simulated power curves to the analytic bounds given $P(Z_i=1) = 0.25$, similar to the results shown earlier in Figure \ref{fig:simsbounds}. Appendix E demonstrates that the analytic bounds derived for the general case where $P(Z_i=1) = p_z$ perform as well as the bounds derived for the special case of $P(Z_i=1) = 0.5$.

\section{Overview and the Method in Context}

Table \ref{tab:recs} presents a summary of the main results for the LATE power analysis introduced in this study, providing the recommended formulas under the various scenarios considered. The formulas presume the use of the Wald IV estimator to test the null hypothesis that the LATE equals $0$ with a two-sided alternative. Recall that the formulas were derived, without loss of generality, under the assumption that $\kappa > 0$.\footnote{If the effect is expected to have a negative value, researchers can simply treat $\kappa$ as the absolute value of the effect size and continue using the same formulas.} The formulas in Table \ref{tab:recs} provide the conservative values for each quantity of interest depending upon whether the probability of treatment assignment is equal or unequal and whether the ordered means assumption is met or not. This includes conservative values for the minimum detectable effect size ($\tilde\kappa$), required sample size ($\tilde N$), and the power ($\widetilde{1 - \beta}$), all computed as a function of the other parameters, with $M = \Phi^{-1}\left(1-\frac{\alpha}{2}\right) + \Phi^{-1}\left(1 - \beta \right)$ and $c^* = \Phi^{-1}(1-\frac{\alpha}{2})$. 

In sum, to perform a conservative power analysis for the LATE, researchers should first identify which of the four cells in Table \ref{tab:recs} best characterizes their particular study context. They can then compute their quantity of interest (e.g. required sample size) based on hypothetical values of the other parameters using the formulas provided in the table. 

If uncertain whether or not the ordered means assumption is likely to be met, it is recommended that researchers operate as if the assumption is \emph{not} met so as to err on the side of conservatism. Refer to the earlier discussion on the factors to consider when assessing the plausibility of the ordered means assumption. Also recall that in the case where noncompliance is one-sided or almost one-sided (i.e. no/few always-takers or no/few never-takers), the ordered means assumption can be simplified to $\bar{Y}_{NT} \leq \bar{Y}_{C}$ or $\bar{Y}_{C} \leq \bar{Y}_{AT}$ as long as the sparse principal stratum is deemed sufficiently small. In addition, researchers should refer to the earlier discussion on how effect sizes may be mapped to absolute effects. Finally, researchers should also note that corner cases exist whereby negative or non-real numbers may be computed for $\tilde\kappa$, which as a practical matter correspond to prohibitively large effect sizes. Hence, if a negative or non-real number is computed for $\tilde\kappa$, researchers should conclude that it will be effectively impossible to detect the effect in the scenario under consideration.

\begin{table}[ht!]
\small
\centering
\caption{Summary of Results: Conservative Formulas for LATE Power Analysis}
\begin{tabular}{c||c|c}
 & $P(Z=1) = 0.5$ & $P(Z=1) \neq 0.5$ \\
 & Equal Assignment Probability & Unequal Assignment Probability \\
 \midrule
 \midrule
  & & \\
$\neg (\bar{Y}_{NT} \leq \bar{Y}_{C} \leq \bar{Y}_{AT})$ & $\tilde\kappa := \frac{2M}{\pi \sqrt{N} - 2M \sqrt{(0.5 - \frac{\pi}{2})(0.5 + \frac{\pi}{2})}}$ & $\tilde\kappa := \frac{2M}{2 \pi \sqrt{N p_z (1-p_z)} - M}$ \\
 & & \\
Ordered Means & $\tilde N := \frac{4 M^2 \left( 1 + \kappa \sqrt{(0.5 - \frac{\pi}{2})(0.5 + \frac{\pi}{2})} \right)^2 }{\kappa^2 \pi^2}$ & $\tilde N := \frac{M^2 (1 + 0.5 \kappa)^2}{p_z (1-p_z) \kappa^2 \pi^2}$ \\
Not Met  & & \\
%  & $\frac{\tau}{\sqrt{V_N^{\hat{\tau}}}} := \sqrt{ \frac{0.25 \kappa^2 N \pi^2}{ 1 + \kappa^2 (0.5 - \frac{\pi}{2})(0.5 + \frac{\pi}{2}) +  2 \kappa  \sqrt{(0.5 - \frac{\pi}{2})(0.5 + \frac{\pi}{2})} } }$ & $\frac{\tau}{\sqrt{V_N^{\hat{\tau}}}} := \sqrt{ \frac{p_z (1-p_z) \kappa^2 N \pi^2}{ 1 + 0.25 \kappa^2 +  \kappa }}$ \\
  & $\widetilde{1 - \beta} :=$ & $\widetilde{1 - \beta} :=$ \\
  &  &  \\
  & $\Phi \left(-c^* + \frac{0.5 \kappa \pi \sqrt{N}}{ 1 + \kappa \sqrt{(0.5 - \frac{\pi}{2})(0.5 + \frac{\pi}{2})} } \right) + $ & $\Phi \left(-c^* + \frac{\kappa \pi \sqrt{p_z (1-p_z) N}}{1 + 0.5 \kappa} \right) + $ \\
  &  &  \\
  & $\Phi \left(-c^* - \frac{0.5 \kappa \pi \sqrt{N}}{ 1 + \kappa \sqrt{(0.5 - \frac{\pi}{2})(0.5 + \frac{\pi}{2})} } \right)$ & $\Phi \left(-c^* - \frac{\kappa \pi \sqrt{p_z (1-p_z) N}}{1 + 0.5 \kappa} \right)$ \\
  & & \\
  & & \\
\midrule
  & & \\
$\bar{Y}_{NT} \leq \bar{Y}_{C} \leq \bar{Y}_{AT}$ & $\tilde\kappa := \frac{2 M}{\sqrt{N \pi^2 - 4 M^2 (0.5 - \frac{\pi}{2})(0.5 + \frac{\pi}{2})}}$ & $\tilde\kappa := \frac{2M}{\sqrt{4 \pi^2 N p_z (1-p_z) - M^2}}$ \\
 & & \\
Ordered Means & $\tilde N :=  \frac{4 M^2 ( 1 + \kappa^2 (0.5 - \frac{\pi}{2})(0.5 + \frac{\pi}{2}))}{\kappa^2 \pi^2}$ & $\tilde N := \frac{M^2 (1 + 0.25 \kappa^2)}{p_z (1-p_z) \kappa^2 \pi^2}$ \\
Met & &  \\
%  & $\frac{\tau}{\sqrt{V_N^{\hat{\tau}}}} := \sqrt{ \frac{0.25 \kappa^2 N \pi^2}{ 1 + \kappa^2 (0.5 - \frac{\pi}{2})(0.5 + \frac{\pi}{2}) } }$ & $\frac{\tau}{\sqrt{V_N^{\hat{\tau}}}} := \sqrt{ \frac{p_z (1-p_z) \kappa^2 N \pi^2}{ 1 + 0.25 \kappa^2 }}$ \\
  & $\widetilde{1 - \beta} :=$ & $\widetilde{1 - \beta} :=$ \\
  &  &  \\ 
  & $\Phi \left(-c^* + \frac{0.5 \kappa \pi \sqrt{N}}{\sqrt{ 1 + \kappa^2 (0.5 - \frac{\pi}{2})(0.5 + \frac{\pi}{2}) }} \right) + $ & $\Phi \left(-c^* + \frac{\kappa \pi \sqrt{p_z (1-p_z)N}}{\sqrt{1 + 0.25 \kappa^2}} \right) + $ \\
  &  &  \\ 
  & $\Phi \left(-c^* - \frac{0.5 \kappa \pi \sqrt{N}}{\sqrt{ 1 + \kappa^2 (0.5 - \frac{\pi}{2})(0.5 + \frac{\pi}{2}) }} \right)$ & $\Phi \left(-c^* - \frac{\kappa \pi \sqrt{p_z (1-p_z)N}}{\sqrt{1 + 0.25 \kappa^2}} \right)$ \\
\end{tabular}
\label{tab:recs}
\end{table}

To illustrate how the method of LATE power analysis presented in this study could be used, the method is applied to the context of the National JTPA Study \citep{bloometal1997}. As described earlier, subjects were randomly assigned such that they were either allowed to enroll in a job training program (assigned to treatment) or excluded from the training for an 18-month period (assigned to control). However, many subjects exhibited noncompliance: many assigned to the treatment decided not to enroll in the training program, while a few assigned to the control gained access to the job training program. The outcome of interest here is the subjects' earnings in the 30-month period following their random assignment.

For the purposes of this illustration, two different values of $\pi$ will be employed. The first is its estimated value as observed in the JTPA data,\footnote{The dataset used here is the tabulation of the JTPA study data by \cite{abadieetal2002}. The data correspond to adult participants in the JTPA experiment for whom 30-month earnings were measured.} which is $0.63$. This is, of course, not necessarily something the researcher would know precisely in advance, but it provides a useful point of reference. The second value for $\pi$ will be $0.4$, which we may view as a researcher's conservative guess prior to the actual study. We will fix $p_Z$ at its observed value of $0.67$, since this is a value over which the researcher has control, and hence the formulas in the far right column of Table \ref{tab:recs} are applicable. We set $\alpha$ and $\beta$ at their conventional levels of $0.05$ and $0.2$, respectively. We can then specify a range of effect sizes ($\kappa$'s) to determine the conservative sample size required ($\tilde{N}$ in Table \ref{tab:recs}) under these specifications. Furthermore, because we know in retrospect the pooled within-assignment-group variance of the outcome (earnings), we can map the $\kappa$ values to absolute effect values ($\tau$'s). Note also that in the JTPA experiment this value is virtually identical to $\widehat{Var}(Y_i|Z_i=0)$, which could have been estimated in advance of the study via a baseline survey given that assignment to control represents a natural state of the world in the absence of intervention.\footnote{$\widehat{\sqrt{E[Var(Y_i|Z_i)]}} = 16759$, while $\widehat{\sqrt{Var(Y_i|Z_i=0)}} = 16180$, a difference of about 3\%.} As a result, the $\kappa$ values could have been mapped to $\tau$ values in the absence of retrospective data, to the benefit of implementing the power analysis.

\begin{table}[ht!]
\footnotesize
\centering
\caption{LATE Power Analysis, Given $\pi = 0.63$ and $p_Z = 0.67$}
\begin{tabular}{p{1.25cm}p{1.9cm}cc}
\toprule
$\kappa$ & $\tau$ & Recommended $\tilde{N}$ without & Recommended $\tilde{N}$ with  \\
         &        & Ordered Means Assumption  & Ordered Means Assumption   \\
\midrule
0.05       & 837.94        & 37588      & 35799     \\
0.10       & 1675.89       & 9861       & 8966      \\
0.15       & 2513.83       & 4594       & 3998      \\
0.20       & 3351.78       & 2706       & 2258      \\
0.25       & 4189.72       & 1811       & 1453      \\
0.30       & 5027.66       & 1314       & 1016      \\
0.35       & 5865.61       & 1008       & 752       \\
0.40       & 6703.55       & 805        & 581       \\
0.45       & 7541.50       & 663        & 464       \\
0.50       & 8379.44       & 559        & 380       \\              
\bottomrule
\end{tabular}
\label{tab:comparison1}
\end{table}

The results given $\pi = 0.63$ are shown in Table \ref{tab:comparison1}, with the conservative recommendation for the required sample size, $\tilde{N}$, provided with and without making Assumption \ref{assump:om} (ordered means). For instance, given a desired effect size of $0.1$, the conservative sample size recommendation would be approximately $10,000$ observations to achieve a level of power of $0.8$ to reject the null hypothesis that $\tau = 0$ without making Assumption \ref{assump:om}, while it would be approximately $9,000$ observations given Assumption \ref{assump:om}. The actual LATE effect size estimate in the pooled adult sample was $0.11$.\footnote{The estimate of the LATE of the training program on earnings is $ \$1,849$. This divided by the expected within-assignment-group standard deviation of earnings in the sample, $16,759$, yields an effect size of 0.11.} Thus, it is no surprise that given the actual sample size of $11,204$ adult participants and the fact that the ordered means assumption was ultimately met in this study, the LATE estimate in the study is statistically significant ($p < 0.001$). While $0.11$ would generally be considered a relatively small effect size---according to the rough guidance presented by \cite{cohen1988}, $0.2$, $0.5$, and $0.8$ are benchmarks for small, medium, and large effect sizes in the social and behavioral sciences---the JTPA study was of sufficiently large scale to detect this effect in the pooled adult sample.

\begin{table}[ht!]
\footnotesize
\centering
\caption{LATE Power Analysis, Given $\pi = 0.4$ and $p_Z = 0.67$}
\begin{tabular}{p{1.25cm}p{1.9cm}cc}
\toprule
$\kappa$ & $\tau$ & Recommended $\tilde{N}$ without & Recommended $\tilde{N}$ with  \\
         &        & Ordered Means Assumption  & Ordered Means Assumption   \\
\midrule
0.05       & 837.94        & 93241      & 88804     \\
0.10       & 1675.89       & 24461      & 22242     \\
0.15       & 2513.83       & 11395      & 9916      \\
0.20       & 3351.78       & 6712       & 5602      \\
0.25       & 4189.72       & 4493       & 3605      \\
0.30       & 5027.66       & 3260       & 2521      \\
0.35       & 5865.61       & 2501       & 1867      \\
0.40       & 6703.55       & 1997       & 1442      \\
0.45       & 7541.50       & 1644       & 1151      \\
0.50       & 8379.44       & 1387       & 943       \\              
\bottomrule
\end{tabular}
\label{tab:comparison2}
\end{table}

Table \ref{tab:comparison2} displays the results given $\pi = 0.4$. As shown, an increase in the amount of noncompliance leads to a disproportionately large increase in the sample size requirements. While noncompliance is assumed to increase by a factor of about $1.6$, the required sample size given any particular $\kappa$ increases by a factor of about $2.5$. As these results show, had the compliance rate $\pi$ actually been 0.4, it is likely that the JTPA study would have failed to find a statistically significant effect of the training program on earnings, even in the pooled adult sample. The method of LATE power analysis presented in this study is designed to alert researchers to such possibilities of under-powered designs before studies are launched without requiring researchers to make the collection of strong assumptions involved in other approaches to LATE power analysis.

\section{Power with Covariates}

The standard LATE assumptions establish the consistency of the Wald IV estimator without covariate adjustment, but covariates can still be used to improve the precision of the estimates. As a result, researchers sometimes employ covariate adjustment in order to attain a more powerful LATE estimator. A common approach is to use linear two-stage least squares (2SLS), which is equivalent to modeling and estimating linear first-stage and intent-to-treat relationships \citep[][pp. 120-122]{angristpischke2009a}:
\begin{eqnarray}
D_i = \mathbf{W}_i \mathbf{\eta} + \pi Z_i + \nu_i^* \label{eq1cov}  \\
Y_i = \mathbf{W}_i \mathbf{\xi} +  \gamma Z_i + \zeta_i^* \label{eq3cov} 
\end{eqnarray}
where $\mathbf{W}_i$ corresponds to a set of covariates, as well as an intercept. Provided that the covariates contained in $\mathbf{W}_i$ are pre-treatment-assignment covariates---that is, they are independent of $Z_i$ and hence do not result in biased estimates of $\pi$ and $\gamma$---then the LATE can be estimated consistently by $\frac{\hat{\gamma}}{\hat{\pi}}$, where $\hat{\gamma}$ and $\hat{\pi}$ are the linear least squares estimators. In addition, if $\mathbf{W}$ helps to explain variation in $D$ and/or $Y$ that is left unexplained by $Z$, then the covariate adjustment can also decrease the variance of $\frac{\hat{\gamma}}{\hat{\pi}}$. As a result, linear 2SLS with covariate adjustment has the potential to offer a more powerful estimator of the LATE, and the method presented in this study can be extended to incorporate these gains. 

\vspace{.1cm}
\begin{definition} \label{R2defs}
Define the following:
$$R^2_{DW} = \frac{\sigma^2 - \sigma^{*2}}{\sigma^2} \:\:\:\:\:\:\:\:\:\: and \:\:\:\:\:\:\:\:\:\: R^2_{YW} = \frac{\omega^2 - \omega^{*2}}{\omega^2}$$
where $\sigma^{2} = E[\nu_i^{2}]$ as defined in the proof of Proposition \ref{step6}, $\omega^{2} = E[\zeta_i^{2}]$ as defined in the proof of Proposition \ref{step6}, $\sigma^{*2} = E[\nu_i^{*2}]$ from equation (\ref{eq1cov}), and $\omega^{*2} = E[\zeta_i^{*2}]$ from equation (\ref{eq3cov}).
\end{definition}
\vspace{.1cm}

$R^2_{DW}$ measures the proportion of variation in $D$ left unexplained by $Z$ that is explained by the covariates contained in $\mathbf{W}$, while $R^2_{YW}$ measures the proportion of variation in $Y$ left unexplained by $Z$ that is explained by the covariates contained in $\mathbf{W}$. Given Definition \ref{R2defs}, covariate adjustment in the 2SLS framework can be employed to yield the following bounds for use in the power formula (continuing to assume that $\kappa > 0$ and $\tau > 0$ are under investigation):
$$\frac{0.5 \kappa \pi \sqrt{N}}{\sqrt{(1-R^2_{YW}) + \kappa^2 (1-R^2_{DW}) E[\nu_i^2] + 2 \kappa \sqrt{(1-R^2_{YW})(1-R^2_{DW}) E[\nu_i^2]} }} $$
$$\leq \frac{\tau}{\sqrt{V_N^{\widehat{2SLS}}}}$$
$$\leq \frac{0.5 \kappa \pi \sqrt{N}}{\sqrt{(1-R^2_{YW}) + \kappa^2 (1-R^2_{DW}) E[\nu_i^2] - 2 \kappa \sqrt{(1-R^2_{YW})(1-R^2_{DW}) E[\nu_i^2]} }} $$

As previously, this formula can be modified to both relax Assumption \ref{assump:epa} (equal assignment probability) and employ Assumption \ref{assump:om} (ordered means), and $E[\nu_i^2]$ replaced with either $(0.5 - \frac{\pi}{2})(0.5 + \frac{\pi}{2})$ or $0.25$ depending on whether Assumption \ref{assump:epa} is made. More detail is provided in the SM Appendix F. It must be emphasized that the results described in this section only apply given the standard LATE assumptions (\ref{assump:sutva}-\ref{assump:m}) as well as independence between $Z$ and $\mathbf{W}$. In other words, the assumptions necessary for the consistency of the estimator must be met without covariate adjustment, the purpose of covariate adjustment must simply be to decrease the variance of the estimator, and the covariates must not be affected by $Z$.

\section{Power with Variable Treatments}

In cases where the endogenous treatment is no longer binary but rather has variable intensity (e.g.  drug dosage, years of schooling), \cite{angristimbens1995} have shown that the Wald IV estimator can still be used under Assumptions \ref{assump:sutva}-\ref{assump:m}. In this case, however, the Wald IV estimator is consistent for a new estimand they call the average causal response (ACR), which is ``a weighted average of causal responses to a unit change in treatment, for those whose treatment status is affected by the instrument" (p. 435). In other words, like with the LATE, the estimand only pertains to those subjects for whom the instrument has a non-zero effect on treatment uptake/dosage, but the ACR is a weighted average rather than a simple average of the individual-level causal effects of the treatment on the outcome.\footnote{See \cite{angristimbens1995} for more details on the weighting formula.}

In spite of the modified estimand, the general properties of the Wald IV estimator, including its variance, remain the same. Furthermore, the assumption of a binary treatment is not critical in the derivation of the power formulas introduced in this study. The binary treatment assumption was employed in determining values for $E[\nu_i^2]$, but a linear re-scaling of a multi-valued treatment to the interval $[0,1]$ would mean the conservative value of $E[\nu_i^2] = 0.25$ would remain valid. As arbitrary linear transformations of variables do not affect statistical power, the method of power analysis presented in this study can also be applied to variable treatments.\footnote{Intuitively, this re-scaling would not affect the power, even though it would mean a re-scaling of $E[\nu_i^2]$, because it would result in a commensurate re-scaling of $\pi$.} Yet the researcher must keep in mind that given a variable treatment, the estimand that is identified is the ACR rather than the LATE, and $\pi$ can no longer be interpreted simply as the compliance rate.

\section{Conclusion}

This study proposed a new approach to power analysis in the LATE context that makes three important contributions. First, in contrast to previous approaches, it does not involve distributional assumptions about the various principal strata. Second, and most importantly, it provides a tight lower bound on the power while removing the need to specify or make assumptions about variance components or distributional heterogeneity across the principal strata. Third, it shows how additional assumptions can be made to raise the lower bound to better balance conservatism with efficiency.

By providing bounds on the power that are free of distributional assumptions, this study introduces a reliable and disciplined way of computing power conservatively without the inefficiencies of other approaches (e.g. setting arbitrarily high variances) that can lead to excessively conservative calculations. The result is a generalized approach to power analysis in the LATE context that is simultaneously conservative, disciplined, and simple to implement. 
%This is an advance over previous power analysis methods in the LATE context, which require making explicit guesses or implicit assumptions about reasonable treatment effect magnitudes, multiple principal strata mean and variance components, and the proportions of the population belonging to each principal stratum.

\addtolength{\baselineskip}{-0.1\baselineskip}
\bibliographystyle{apalike}
\bibliography{references}

\clearpage

\renewcommand{\thepage}{\roman{page}}
\setcounter{page}{0}
\setcounter{footnote}{0}

\Large
\begin{center}
\textbf{Supplementary Materials} \\ 
\vspace{1cm}
for \\
\vspace{4cm}
\LARGE
A Generalized Approach to Power Analysis \\ for Local Average Treatment Effects \\
\vspace{1cm}
\Large
Kirk Bansak \\
\vspace{10cm}
\normalsize
\monthname \ \number\year
\end{center}

\normalsize

\clearpage
\addtolength{\baselineskip}{0.2\baselineskip}

\section*{Appendix A: Proofs}

\setcounter{table}{0}
\renewcommand{\thetable}{A\arabic{table}}%

\vspace{1cm}

\noindent \emph{Proof of Proposition \ref{step6}.} \\

The asymptotic variance of the Wald IV estimator, as shown by \cite{imbensangrist1994}, is:
$$V_N^{\hat{\tau}} = \frac{E[\epsilon_i^2 \{ Z_i - E[Z_i] \}^2]}{N Cov^2(D_i,Z_i)}$$
where $\epsilon_i = Y_i - E[Y_i] - \tau(D_i - E[D_i])$. This variance can be simplified as follows:

\vspace{1cm}
\begin{lemma} \label{step1}
Given Assumption \ref{assump:epa} (equal assignment probability),\footnote{Proofs of all lemmas can be found below.} 
$$V_N^{\hat{\tau}} = \frac{E[\epsilon_i^2]}{0.25 N \pi^2} $$
where $\epsilon_i = Y_i - E[Y_i] - \tau(D_i - E[D_i])$.
\end{lemma}
\vspace{1cm}

Letting $\zeta_i = Y_i - E[Y_i] - \gamma(Z_i - E[Z_i])$ and $\nu_i = D_i - E[D_i] - \pi(Z_i - E[Z_i])$, the variance can then be re-expressed as follows:

\vspace{1cm}
\begin{lemma} \label{step2}
Given Assumption \ref{assump:epa} (equal assignment probability), $$V_N^{\hat{\tau}} = \frac{E[\zeta_i^2] + \tau^2 E[\nu_i^2] - 2 \tau E[\zeta_i \nu_i]}{0.25 N \pi^2}$$
where $\zeta_i = Y_i - E[Y_i] - \gamma(Z_i - E[Z_i])$ and $\nu_i = D_i - E[D_i] - \pi(Z_i - E[Z_i])$.
\end{lemma}
\vspace{1cm}

%The value and even sign of $E[\zeta_i \nu_i]$ cannot be known without making additional assumptions. However, $E[\zeta_i \nu_i]$ can be eliminated by employing the following strict bounds:
These tight bounds then follow:

\vspace{1cm}
\begin{lemma} \label{step3}
Assuming without loss of generality that $\tau > 0$, $$\frac{E[\zeta_i^2] + \tau^2 E[\nu_i^2] - 2 \tau \sqrt{E[\nu_i^2]} \sqrt{E[\zeta_i^2]}}{0.25 N \pi^2} \leq V_N^{\hat{\tau}} \leq \frac{E[\zeta_i^2] + \tau^2 E[\nu_i^2] + 2 \tau \sqrt{E[\nu_i^2]} \sqrt{E[\zeta_i^2]}}{0.25 N \pi^2}$$
\end{lemma}
\vspace{1cm}

%Hence, assuming without loss of generality that $\tau > 0$, we have the following bounds for $V_N^{\hat{\tau}}$:
%$$ \frac{\sigma_3^2 + \tau^2 \sigma_1^2 \pm 2 \tau \sigma_1 \sigma_3}{0.25 N \pi^2} $$
%Alternatively, given that $\tau = \frac{\gamma}{\pi}$, the bounds can also be expressed as:
%$$\frac{E[\zeta_i^2] + \tau^2 E[\nu_i^2] - 2 \tau \sqrt{E[\nu_i^2]} \sqrt{E[\zeta_i^2]}}{0.25 N \pi^2} \leq V_N^{\hat{\tau}} \leq \frac{E[\zeta_i^2] + \tau^2 E[\nu_i^2] + 2 \tau \sqrt{E[\nu_i^2]} \sqrt{E[\zeta_i^2]}}{0.25 N \pi^2}$$

%And what's really nice, as you can see, $- 2 \frac{\gamma}{\pi} \sigma_1 \sigma_3$ will have a $\sigma_3^2$ in it, and so everything will continue to have a $\sigma_3^2$ in it, allowing us to factor it out as before, thus eliminating all outcome variable variances in both the conservative and liberal versions! Thus, in the end, we can bound the power!!!

%\subsection{BACK TO BEFORE!!!}

As established by Definition \ref{step4}, let $\kappa = \frac{\tau}{\sqrt{E[Var(Y_i|Z_i)]}}$. It is then possible re-express $\tau$ as follows:

\vspace{1cm}
\begin{lemma} \label{step5}
Given Definition \ref{step4},
$$\tau = \kappa \sqrt{E[\zeta_i^2]}$$
%and
%$$\gamma = \sqrt{\frac{\kappa^2 \sigma_3^2 \pi^2}{1 - 0.25 \kappa^2 \pi^2}}$$
\end{lemma}
\vspace{1cm}

Note that, without loss of generality, it is assumed that $\tau > 0$, and for convenience, consider the square of $\frac{\tau}{\sqrt{V_N^{\hat{\tau}}}}$. Given Assumption \ref{assump:epa} (equal assignment probability), Definition \ref{step4}, and Lemmas \ref{step1}-\ref{step5}, the bounds on $\frac{\tau^2}{V_N^{\hat{\tau}}}$ (expressed using the $\pm$ operator) are the following:

\begin{eqnarray}
\nonumber  \frac{0.25 \kappa^2 E[\zeta_i^2] N \pi^2}{E[\zeta_i^2] + \tau^2 E[\nu_i^2] \pm 2 \tau \sqrt{E[\zeta_i^2]}\sqrt{E[\nu_i^2]}} \\
\nonumber = \frac{0.25 \kappa^2 E[\zeta_i^2] N \pi^2}{E[\zeta_i^2] + \kappa^2 E[\zeta_i^2] E[\nu_i^2] \pm 2 \kappa E[\zeta_i^2]\sqrt{E[\nu_i^2]}} \\
\nonumber = \frac{0.25 \kappa^2 \pi^2 N}{1 + \kappa^2 E[\nu_i^2] \pm 2 \kappa \sqrt{E[\nu_i^2]}} \label{eq:postcancel}
\end{eqnarray}

%Notably, $E[\zeta_i^2]$ factors out of both the numerator and denominator. 
This yields the following bounds on $\frac{\tau}{\sqrt{V_N^{\hat{\tau}}}}$:
$$\sqrt{ \frac{0.25 \kappa^2 \pi^2 N}{ 1 + \kappa^2 E[\nu_i^2] +  2 \kappa  \sqrt{E[\nu_i^2]} } } \leq \frac{\tau}{\sqrt{V_N^{\hat{\tau}}}} \leq \sqrt{ \frac{0.25 \kappa^2 \pi^2 N}{ 1 + \kappa^2 E[\nu_i^2] -  2 \kappa  \sqrt{E[\nu_i^2]} } }$$

$\square$

\vspace{2cm}
\clearpage

\emph{Proof of Lemma \ref{step1}.} \\

%The asymptotic variance of the Wald IV estimator, as shown by \cite{imbensangrist1994}, is:
%$$V_N^{\hat{\tau}} = \frac{E[\epsilon_i^2 \{ Z_i - E[Z_i] \}^2]}{N Cov^2(D_i,Z_i)}$$
%where $\epsilon_i = Y_i - E[Y_i] - \tau(D_i - E[D_i])$. \\

Given Assumption \ref{assump:epa} (equal assignment probability): \\ 
$$E[\epsilon_i^2 \{ Z_i - E[Z_i] \}^2] = E[\epsilon_i^2 \{ Z_i - 0.5 \}^2] = E[E[\epsilon_i^2 \{ Z_i - 0.5 \}^2 | Z_i]]$$
%$$ = p(Z_i=0)E[\epsilon_i^2 \{ 0 - 0.5 \}^2 | Z_i=0] + p(Z_i=1)E[\epsilon_i^2 \{ 1 - 0.5 \}^2 | Z_i=1]$$
$$ = p(Z_i=0)E[\epsilon_i^2 \{ 0 - 0.5 \}^2] + p(Z_i=1)E[\epsilon_i^2 \{ 1 - 0.5 \}^2]$$
%$$ = 0.5 E[\epsilon_i^2 0.25 | Z_i=0] + 0.5E[\epsilon_i^2 0.25 | Z_i=1]$$ 
%$$ = \{0.5 E[\epsilon_i^2|Z_i=0] + 0.5E[\epsilon_i^2|Z_i=1] \} 0.25$$
$$ = E[\epsilon_i^2] 0.25$$

Thus, $$V_N^{\hat{\tau}} = \frac{E[\epsilon_i^2 \{ Z_i - E[Z_i] \}^2]}{N Cov^2(D_i,Z_i)} = \frac{E[\epsilon_i^2] Var(Z_i)}{N Cov^2(D_i,Z_i)}$$

Finally, since $\frac{Cov(D_i,Z_i)}{Var(Z_i)} = \pi$ and $Var(Z_i) = 0.25$, this further simplifies to:
%Further, this can be re-expressed as:
$$V_N^{\hat{\tau}} = \frac{E[\epsilon_i^2]}{N} \frac{Var(Z_i)}{Cov^2(D_i,Z_i)} \frac{Var(Z_i)}{Var(Z_i)} = \frac{E[\epsilon_i^2]}{N} \left(\frac{Var(Z_i)}{Cov(D_i,Z_i)} \right)^2 \frac{1}{Var(Z_i)} = \frac{E[\epsilon_i^2]}{0.25 N \pi^2} $$
$\square$

\vspace{2cm}
\clearpage

\noindent \emph{Proof of Lemma \ref{step2}.} \\

Let $\zeta_i = Y_i - E[Y_i] - \gamma(Z_i - E[Z_i])$ and $\nu_i = D_i - E[D_i] - \pi(Z_i - E[Z_i])$. Given $\epsilon_i = Y_i - E[Y_i] - \tau(D_i - E[D_i])$ and $\frac{\gamma}{\pi} = \tau$ under Assumptions \ref{assump:sutva}-\ref{assump:m}:
$$\epsilon_i = \zeta_i - \tau \nu_i$$
Hence,
$$E[\epsilon_i^2] = E[(\zeta_i - \tau \nu_i)^2] = E[\zeta_i^2] + \tau^2 E[\nu_i^2] - 2 \tau E[\nu_i \zeta_i]$$
%where $\sigma_1^2 = Var(\nu_i)$, $\sigma_3^2 = Var(\zeta_i)$, and $\sigma_{13} = Cov(\nu_i,\zeta_i)$, since $E[\nu_i] = E[\zeta_i] = 0$. 

Combined with Lemma \ref{step1}, this results in:
$$V_N^{\hat{\tau}} = \frac{E[\zeta_i^2] + \tau^2 E[\nu_i^2] - 2 \tau E[\zeta_i \nu_i]}{0.25 N \pi^2} $$
$\square$

\vspace{2cm}
\clearpage

\noindent \emph{Proof of Lemma \ref{step3}.} \\

Given $E[\zeta_i] = E[\nu_i] = 0$, $E[\zeta_i \nu_i] = Cov(\zeta_i,\nu_i)$. By the Cauchy-Schwarz inequality, $E[\zeta_i\nu_i]$ is bounded as follows:
$$- \sqrt{E[\zeta_i^2]E[\nu_i^2]} = - \sqrt{ Var(\nu_i) Var(\zeta_i) } \leq E[\zeta_i \nu_i] = Cov(\nu_i,\zeta_i)$$
$$\leq \sqrt{ Var(\nu_i) Var(\zeta_i) } = \sqrt{E[\zeta_i^2]E[\nu_i^2]}$$

Hence, assuming without loss of generality that $\tau>0$, $E[\zeta_i\nu_i]$ can be set to its lower-bound value (thereby setting $-2 \tau E[\zeta_i \nu_i]$ to its upper-bound value), yielding the following:
$$E[\zeta_i^2] + \tau^2 E[\nu_i^2] - 2 \tau E[\zeta_i \nu_i] \leq 
E[\zeta_i^2] + \tau^2 E[\nu_i^2] - 2 \tau (- \sqrt{E[\zeta_i^2]E[\nu_i^2]})$$ 
$$= E[\zeta_i^2] + \tau^2 E[\nu_i^2] + 2 \tau \sqrt{E[\zeta_i^2]E[\nu_i^2]}$$
As a result, this component in the denominator of $\frac{\tau}{V_N^{\hat{\tau}}}$ is set to its largest possible value, making $\frac{\tau}{V_N^{\hat{\tau}}}$ as small as possible. This will yield a tight lower bound on the power formula. A tight upper bound on the power can also be computed by setting $E[\zeta_i\nu_i]$ to its upper-bound value.
$\square$

\vspace{2cm}
\clearpage

\noindent \emph{Proof of Lemma \ref{step5}.} \\

Given $\zeta_i = Y_i - E[Y_i] - \gamma(Z_i - E[Z_i])$, we have that $Y_i = \zeta_i + E[Y_i] + \gamma(Z_i - E[Z_i])$. Hence,
$$Var(Y_i|Z_i) = Var(\zeta_i + E[Y_i] + \gamma(Z_i - E[Z_i]) |Z_i) = Var(\zeta_i|Z_i)$$
As a result, 
$$E[Var(Y_i|Z_i)] = E[Var(\zeta_i|Z_i)] = Var(\zeta_i) - Var(E[\zeta_i|Z_i])$$
by the law of total variance. Further, since $\gamma = E[Y_i|Z_i=1] - E[Y_i|Z_i=0]$ by assumptions \ref{assump:sutva}-\ref{assump:m}, it can be shown that $E[\zeta_i|Z_i] = E[\zeta_i|Z_i=0] = E[\zeta_i|Z_i=1] = 0$. For $E[\zeta_i|Z_i=0]$:
$$E[\zeta_i|Z_i=0] = E[Y_i - E[Y_i] - \gamma(Z_i - E[Z_i])|Z_i=0] = E[Y_i|Z_i=0] - E[Y_i] + \gamma E[Z_i]$$
$$=E[Y_i|Z_i=0] - \{ E[Y_i|Z_i=0](1-E[Z_i]) + E[Y_i|Z_i=1]E[Z_i] \}$$ 
$$+ \{ E[Y_i|Z_i=1]E[Z_i] - E[Y_i|Z_i=0]E[Z_i] \}$$
$$=0$$
The result follows similarly for $E[\zeta_i|Z_i=1]$. Therefore,
$$E[Var(Y_i|Z_i)] = E[Var(\zeta_i|Z_i)] = Var(\zeta_i) - Var(E[\zeta_i|Z_i]) = Var(\zeta_i) = E[\zeta_i^2]$$
and hence
$$\tau = \kappa \sqrt{E[\zeta_i^2]}$$
$\square$

\vspace{2cm}
\clearpage

\noindent \emph{Proof of Proposition \ref{step8}.} \\

First, the following can be shown.

\vspace{1cm}
\begin{lemma} \label{step7}
$$E[\nu_i^2] \leq \left(0.5 - \frac{\pi}{2} \right) \left(0.5 + \frac{\pi}{2} \right)$$
\end{lemma}
\vspace{1cm}

Then, substitute $(0.5 - \frac{\pi}{2})(0.5 + \frac{\pi}{2})$ in for $E[\nu_i^2]$ in the result in Proposition \ref{step6} (and simplify). 
$\square$

\vspace{2cm}
\clearpage

\noindent \emph{Proof of Lemma \ref{step7}.} \\

Consider that, by Assumption \ref{assump:epa} (equal assignment probability):
$$E[\nu_i^2] = E[Var(D_i|Z_i)] = 0.5 Var(D_i|Z_i=0) + 0.5 Var(D_i|Z_i=1)$$ 
$$= 0.5 [Var(D_i|Z_i=0) + Var(D_i|Z_i=1)]$$
Thus, $E[\nu_i^2]$ is maximized by maximizing $Var(D_i|Z_i=0) + Var(D_i|Z_i=1)$. 

Further, consider that $D$ is binary with $E[D_i|Z_i=1] = E[D_i|Z_i=0] + \pi$. As a result, $Var(D_i|Z_i=0) + Var(D_i|Z_i=1)$ is maximized when $E[D_i|Z_i=1]$ and $E[D_i|Z_i=0]$ are equidistant from $0.5$. This occurs when $E[D_i|Z_i=0] = 0.5 - \frac{\pi}{2}$ and $E[D_i|Z_i=1] = 0.5 + \frac{\pi}{2}$. The result is the following upper-bound value:

$$E[\nu_i^2] = E[Var(D_i|Z_i)] = 0.5 Var(D_i|Z_i=0) + 0.5 Var(D_i|Z_i=1)$$ 
$$= 0.5 \left(0.5 - \frac{\pi}{2} \right) \left(0.5 + \frac{\pi}{2} \right) + 0.5 \left(0.5 + \frac{\pi}{2} \right) \left(0.5 - \frac{\pi}{2} \right) $$
$$= \left(0.5 - \frac{\pi}{2} \right) \left(0.5 + \frac{\pi}{2} \right)$$
$\square$

\vspace{2cm}
\clearpage

\noindent \emph{Proof of Proposition \ref{newbound}.} \\

Given that the power formula bounds were based on the bounds of $E[\nu_i \zeta_i] = Cov(\nu_i,\zeta_i)$, the lower bound can be increased if it is the case that $Cov(\nu_i,\zeta_i) \geq 0$. Assuming without loss of generality that $\tau > 0$, it can be shown that this condition is met when $\bar{Y}_{NT} \leq \bar{Y}_{C} \leq \bar{Y}_{AT}$, where $\bar{Y}_{C}$, $\bar{Y}_{NT}$, and $\bar{Y}_{AT}$ denote the expected realized outcome value for compliers, never-takers, and always-takers (e.g. $\bar{Y}_{C} = E[Y_i|Complier]$).

\vspace{1cm}
\begin{lemma} \label{narrow}
Letting $X_i$ denote the principal stratum to which unit $i$ belongs and assuming without loss of generality that $\tau > 0$,
$$Cov(E[D_i|X_i],E[Y_i|X_i]) \geq 0 \Longrightarrow Cov(\nu_i,\zeta_i) \geq 0$$
$$\therefore \:\: \bar{Y}_{NT} \leq \bar{Y}_{C} \leq \bar{Y}_{AT} \Longrightarrow Cov(\nu_i,\zeta_i) \geq 0$$
\end{lemma}
\vspace{1cm}

Therefore, given Assumption \ref{assump:om}, $Cov(\nu_i,\zeta_i)$ is non-negative and hence the lower bound of the power calculation results from $Cov(\nu_i,\zeta_i) = 0$. 
%By Lemma \ref{narrow}, $\bar{Y}_{NT} \leq \bar{Y}_{C} \leq \bar{Y}_{AT} \Longrightarrow Cov(\nu_i,\zeta_i) \geq 0$. The value that produces the most conservative (i.e. lowest) power calculation is $Cov(\nu_i,\zeta_i) = 0$. 
Given $Cov(\nu_i,\zeta_i) = 0$, then $E[\nu_i \zeta_i] = 0$, and using select results from Proposition 1:
$$\frac{0.25 \kappa^2 \pi^2 N}{ 1 + \kappa^2 E[\nu_i^2] } \leq \frac{\tau^2}{V_N^{\hat{\tau}}}$$
Finally, since $D$ is binary, $E[\nu_i^2]$ can be simplified as a function of $\pi$, with $E[\nu_i^2] = (0.5 - \frac{\pi}{2})(0.5 + \frac{\pi}{2})$ being the most conservative by Lemma \ref{step7}. 

The final result follows:
%$$\sqrt{ \frac{0.25 \kappa^2 \pi^2 N}{ 1 + \kappa^2 (0.5 - \frac{\pi}{2})(0.5 + \frac{\pi}{2}) } } \leq \frac{\tau}{\sqrt{V_N^{\hat{\tau}}}}$$
$$ \frac{0.5 \kappa \pi \sqrt{N}}{\sqrt{ 1 + \kappa^2 (0.5 - \frac{\pi}{2})(0.5 + \frac{\pi}{2}) }}  \leq \frac{\tau}{\sqrt{V_N^{\hat{\tau}}}}$$
$\square$ \\
%Hence, the lower bound for the power becomes the following:
%$$\Phi \left(-c^* + \sqrt{ \frac{0.25 \kappa^2 \pi^2 N}{ 1 + \kappa^2 (0.5 - \frac{\pi}{2})(0.5 + \frac{\pi}{2}) } }\right) + \Phi \left(-c^* - \sqrt{ \frac{0.25 \kappa^2 \pi^2 N}{ 1 + \kappa^2 (0.5 - \frac{\pi}{2})(0.5 + \frac{\pi}{2}) } }\right) \leq 1-\beta$$

\noindent Note: In the case of one-sided noncompliance, the missing principal stratum will not factor into $Cov(E[D_i|X_i],E[Y_i|X_i])$. This simplifies Assumption \ref{assump:om} (ordered means assumption) to $\bar{Y}_{NT} \leq \bar{Y}_{C}$ or $\bar{Y}_{C} \leq \bar{Y}_{AT}$, depending upon the direction of noncompliance. It should also be noted that, in the case where noncompliance is almost one-sided (i.e. very few never-takers or very few always-takers), the sparse principal stratum will have only a negligible impact on $Cov(E[D_i|X_i],E[Y_i|X_i])$. Thus, as a practical matter, the ordered means assumption can be simplified to $\bar{Y}_{NT} \leq \bar{Y}_{C}$ or $\bar{Y}_{C} \leq \bar{Y}_{AT}$ as long as the sparse principal stratum is deemed sufficiently small.

\vspace{2cm}
\clearpage

\noindent \emph{Proof of Lemma \ref{narrow}.} \\

Recall that:
\begin{eqnarray}
\nonumber \nu_i = D_i - E[D_i] - \pi (Z_i + E[Z_i]) \\
\nonumber \zeta_i = Y_i - E[Y_i] - \gamma (Z_i +  E[Z_i])
\end{eqnarray}
Hence
\begin{eqnarray}
\nonumber Cov(\nu_i,\zeta_i) = Cov(D_i - \pi Z_i, Y_i - \gamma Z_i) \\
\nonumber = Cov(D_i,Y_i) - \pi Cov(Z_i,Y_i) - \gamma Cov(D_i,Z_i) + \pi \gamma Var(Z_i) \\
\nonumber = Cov(D_i,Y_i) - \pi Cov(Z_i,Y_i) - \gamma Cov(D_i,Z_i) + \frac{Cov(D_i,Z_i)}{Var(Z_i)} \gamma Var(Z_i) \\
\nonumber = Cov(D_i,Y_i) - \pi Cov(Z_i,Y_i)
\end{eqnarray}

It is further possible to decompose $Cov(D_i,Y_i)$ using the law of total covariance, conditioning on principal strata. Let $X_i$ denote the principal stratum to which unit $i$ belongs, making the usual assumption of no defiers. Continue to assume that $P(Z_i=1) = 0.5$, and WLOG that $\tau > 0$. Let $\bar{Y}_{C}$, $\bar{Y}_{NT}$, and $\bar{Y}_{AT}$ denote the expected realized outcome value for compliers, never-takers, and always-takers (e.g. $\bar{Y}_{C} = E[Y_i|Complier]$). Then, 

\begin{eqnarray}
\nonumber Cov(D_i,Y_i) = E[Cov(D_i,Y_i|X_i)] + Cov(E[D_i|X_i],E[Y_i|X_i]) \\
\nonumber = P(Complier) Cov(D_i,Y_i | Complier) + Cov(E[D_i|X_i],E[Y_i|X_i])
\end{eqnarray}
since $Cov(D_i,Y_i | AlwaysTaker) = Cov(D_i,Y_i | NeverTaker)=0$. 

Furthermore, we can show that:
\begin{eqnarray}
\nonumber P(Complier) Cov(D_i,Y_i | Complier) = P(Complier) Cov(Z_i,Y_i | Complier) \\
\nonumber = \pi \frac{Cov(Z_i,Y_i|Complier)}{Var(Z_i|Complier)}Var(Z_i|Complier) \\
\nonumber = \pi \frac{Cov(D_i,Y_i|Complier)}{Var(D_i|Complier)}Var(Z_i) \\
\nonumber = \pi \tau Var(Z_i) \\
\nonumber = \gamma Var(Z_i) \\
\nonumber = \frac{Cov(Z_i,Y_i)}{Var(Z_i)} Var(Z_i) \\
\nonumber = Cov(Z_i,Y_i)
\end{eqnarray}

Hence:
\begin{eqnarray}
\nonumber Cov(\nu_i,\zeta_i) = Cov(D_i,Y_i) - \pi Cov(Z_i,Y_i) \\
%\nonumber = Cov(Z_i,Y_i | Compliers) + Cov(E[D_i|X],E[Y_i|X_i]) - \pi Cov(Z_i,Y_i) \\
\nonumber = Cov(Z_i,Y_i) + Cov(E[D_i|X_i],E[Y_i|X_i]) - \pi Cov(Z_i,Y_i) \\
\nonumber = (1 - \pi)Cov(Z_i,Y_i) + Cov(E[D_i|X_i],E[Y_i|X_i])
\end{eqnarray}

The first term $(1 - \pi)Cov(Z_i,Y_i)$ is non-negative by the assumption (WLOG) that $\tau > 0$. Hence, $Cov(\nu_i,\zeta_i)$ is guaranteed to be non-negative if $Cov(E[D_i|X_i],E[Y_i|X_i])$ is non-negative, which is true when $\bar{Y}_{NT} \leq \bar{Y}_{C} \leq \bar{Y}_{AT}$. $\square$

\vspace{2cm}
\clearpage

\noindent \emph{Proof of Proposition \ref{relaxedbounds}.} \\

Given $E[\epsilon_i^2 | Z_i] = E[\epsilon_i^2]$, then regardless of the value of $P(Z_i=1)$, the following holds:
$$V_N^{\hat{\tau}} = \frac{E[\epsilon_i^2 \{ Z_i - E[Z_i] \}^2]}{N Cov^2(D_i,Z_i)} = \frac{E[\epsilon_i^2] Var(Z_i)}{N Cov^2(D_i,Z_i)}$$

%As a result, Assumption \ref{assump:epa} (equal assignment probability) is no longer necessary, as it previously was in Lemma \ref{step1}. 
Lemmas \ref{step1}-\ref{narrow} and Propositions \ref{step6}-\ref{newbound} then proceed as before with two exceptions. First, whereas previously $Var(Z_i) = 0.25$, now $Var(Z_i) = p_z (1-p_z)$, where $p_z = P(Z_i=1)$. 
%Hence, $V_N^{\hat{\tau}} = \frac{E[\epsilon_i^2]}{p_z (1-p_z) N \pi^2}$. 
Second, $E[\nu_i^2]$ is set to its upper limiting value of $0.25$. 
%This is conservative because $D$ is binary, which means its maximum marginal variance is $0.25$, a value that $E[\nu_i^2]$ cannot exceed.
$\square$

\vspace{2cm}
\clearpage

\noindent \emph{Proof of Proposition \ref{newnewbound}.} \\

The proof follows from the results of Lemma \ref{narrow} applied to Proposition \ref{relaxedbounds}. $\square$

\clearpage

\section*{Appendix B: LATE Power Simulations}

\setcounter{table}{0}
\renewcommand{\thetable}{B\arabic{table}}%

This appendix presents the results of a series of simulations illustrating the power of the test of the null hypothesis that the LATE equals zero using the Wald IV estimator, as the marginal distributional characteristics of the principal strata and other parameters are varied. Each simulation specifies the data-generating distribution of the quadruples $(Y_i(0),Y_i(1),D_i(0),D_i(1)) \in \mathbb{R} \times \mathbb{R} \times \{0,1\} \times \{0,1\}$, randomly draws from that distribution, and randomizes the treatment assignment variable, generating samples of $N$ independent and identically distributed units of the form $(Y_i, D_i, Z_i) \in \mathbb{R} \times \{0,1\} \times \{0,1\}$.

\begin{table}[ht!]
\footnotesize
\centering
\caption{LATE Power Simulations, Varying the Investigation Parameters}
\begin{tabular}{p{8cm}cc}
\toprule
Fixed Parameters: & Varying Parameter & Power \\
\midrule
$\tau = 5$, $\pi = 0.2$, & $N = $ & \:\:\:\:\:\:\:\:\:\:\:\:\:\:\:\:\:\:\:\:\:\:\:\:\:\:\:\:\:\: \\
$E[Y_C(0)] = 0$, $SD(Y_C(0)) = 8$, $SD(Y_C(1)) = 8$,                & $1000$ & $\bf{0.43}$ \\
$\dot{\pi}_{NT} = 0.4$, $E[Y_{NT}(0)] = -3$, $SD(Y_{NT}(0)) = 12$, & $2000$ & $\bf{0.71}$   \\
$\dot{\pi}_{AT} = 0.4$,  $E[Y_{AT}(1)] = 3$, $SD(Y_{AT}(1)) = 4$   & $4000$ & $\bf{0.93}$   \\
\midrule
$N = 1000$, $\pi = 0.2$, & $\tau = $ & \:\:\:\:\:\:\:\:\:\:\:\:\:\:\:\:\:\:\:\:\:\:\:\:\:\:\:\:\:\: \\
$E[Y_C(0)] = 0$, $SD(Y_C(0)) = 8$, $SD(Y_C(1)) = 8$,                & $5$    & $\bf{0.43}$ \\
$\dot{\pi}_{NT} = 0.4$, $E[Y_{NT}(0)] = -3$, $SD(Y_{NT}(0)) = 12$, & $6$    & $\bf{0.57}$   \\ 
$\dot{\pi}_{AT} = 0.4$,  $E[Y_{AT}(1)] = 3$, $SD(Y_{AT}(1)) = 4$   & $7$    & $\bf{0.68}$   \\
\bottomrule
\end{tabular}
\label{tab:sims1}
\end{table}

In all of the simulations presented, $\alpha$ is held fixed at $0.05$. The sample size $N$, LATE $\tau$, and first-stage effect $\pi$ are specified, and samples are generated with specified probabilities that units drawn are compliers ($\pi$), never-takers ($\dot{\pi}_{NT}$), or always-takers ($\dot{\pi}_{AT}$). For instance, the probability that a unit is a complier is $\pi = P(D_i(1) - D_i(0) = 1)$. 

For convenience, the subscript $C$ is used to denote parameters for compliers, $NT$ for never-takers, and $AT$ for always-takers. For instance, for the compliers, the control condition potential outcome mean $E[Y_i(0) | D_i(1) - D_i(0) = 1]$ is denoted by $E[Y_C(0)]$. This parameter $E[Y_C(0)]$ is specified (the treatment condition potential outcome mean for compliers is determined by the specified $\tau$), as are control condition and treatment condition potential outcome standard deviations $SD(Y_C(0))$ and $SD(Y_C(1))$, with potential outcome values being randomly generated to be normally distributed. The potential outcome distributions of the never-takers and always-takers are also randomly generated to be normally distributed. For the never-takers, only a control condition mean and standard deviation, $E[Y_{NT}(0)]$ and $SD(Y_{NT}(0))$, must be specified, and for the always-takers, only a treatment condition mean and standard deviation, $E[Y_{AT}(1)]$ and $SD(Y_{AT}(1))$, must be specified since only those parameters factor into the realized values of $Y_i$ for never-takers and always-takers respectively. Finally, the treatment is randomly assigned with probability one half: $P(Z_i = 1) = 0.5$. Whether a unit is actually treated, and hence obtains a realized outcome value that is the treatment (or control) condition potential outcome, is determined not only by treatment assignment but also the principal stratum to which that unit belongs.

\begin{table}[ht!]
\footnotesize
\centering
\caption{LATE Power Simulations, Varying the Distribution Parameters}
\begin{tabular}{p{8cm}cc}
\toprule
Fixed Parameters: & Varying Parameter(s) & Power \\
\midrule
$N = 1000$, $\tau = 5$, $\pi = 0.2$, & $E[Y_C(0)] = $ & \:\:\:\:\:\:\:\:\:\:\:\:\:\:\:\:\:\:\:\:\:\:\:\:\:\:\:\:\:\: \\
$SD(Y_C(0)) = 8$, $SD(Y_C(1)) = 8$, & $0$ & $\bf{0.45}$ \\
$\dot{\pi}_{NT} = 0.4$, $E[Y_{NT}(0)] = -3$, $SD(Y_{NT}(0)) = 12$,   & $10$    & $\bf{0.35}$   \\ 
$\dot{\pi}_{AT} = 0.4$,  $E[Y_{AT}(1)] = 3$, $SD(Y_{AT}(1)) = 4$     & $20$    & $\bf{0.25}$   \\
\midrule
$N = 1000$, $\tau = 5$, $\pi = 0.3$, & $SD(Y_C(0)),SD(Y_C(1)) = $ & \:\:\:\:\:\:\:\:\:\:\:\:\:\:\:\:\:\:\:\:\:\:\:\:\:\:\:\:\:\: \\
$E[Y_C(0)] = 0$, & $8,8$ & $\bf{0.76}$ \\
$\dot{\pi}_{NT} = 0.35$, $E[Y_{NT}(0)] = -3$, $SD(Y_{NT}(0)) = 12$,  & $8,16$  & $\bf{0.63}$   \\ 
$\dot{\pi}_{AT} = 0.35$,  $E[Y_{AT}(1)] = 3$, $SD(Y_{AT}(1)) = 4$    & $16,16$ & $\bf{0.52}$   \\
\midrule
$N = 1000$, $\tau = 5$, $\pi = 0.2$, & $E[Y_{NT}(0)], E[Y_{AT}(1)] = $ & \:\:\:\:\:\:\:\:\:\:\:\:\:\:\:\:\:\:\:\:\:\:\:\:\:\:\:\:\:\: \\
$E[Y_C(0)] = 0$, $SD(Y_C(0)) = 8$, $SD(Y_C(1)) = 8$,                  & $-3,3$  & $\bf{0.44}$ \\
$\dot{\pi}_{NT} = 0.4$, $SD(Y_{NT}(0)) = 12$,                       & $10,3$  & $\bf{0.26}$   \\ 
$\dot{\pi}_{AT} = 0.4$, $SD(Y_{AT}(1)) = 4$                         & $10,-6$ & $\bf{0.13}$   \\
\midrule
$N = 1000$, $\tau = 5$, $\pi = 0.2$, & $SD(Y_{NT}(0)), SD(Y_{AT}(1)) = $ & \:\:\:\:\:\:\:\:\:\:\:\:\:\:\:\:\:\:\:\:\:\:\:\:\:\:\:\:\:\: \\
$E[Y_C(0)] = 0$, $SD(Y_C(0)) = 8$, $SD(Y_C(1)) = 8$,                  & $12,4$  & $\bf{0.43}$ \\
$\dot{\pi}_{NT} = 0.4$, $E[Y_{NT}(0)] = -3$,                        & $12,8$  & $\bf{0.37}$   \\ 
$\dot{\pi}_{AT} = 0.4$,  $E[Y_{AT}(1)] = 3$                         & $24,8$  & $\bf{0.15}$   \\
\midrule
$N = 1000$, $\tau = 5$, & $\pi$, $\dot{\pi}_{NT}, \dot{\pi}_{AT} =$ & \:\:\:\:\:\:\:\:\:\:\:\:\:\:\:\:\:\:\:\:\:\:\:\:\:\:\:\:\:\: \\
$E[Y_C(0)] = 0$, $SD(Y_C(0)) = 8$, $SD(Y_C(1)) = 8$,                  & $0.3,0.35,0.35$ & $\bf{0.76}$ \\
$E[Y_{NT}(0)] = -3$, $SD(Y_{NT}(0)) = 12$,                           & $0.2,0.4,0.4$   & $\bf{0.45}$   \\ 
$E[Y_{AT}(1)] = 3$, $SD(Y_{AT}(1)) = 4$                              & $0.2,0.1,0.7$   & $\bf{0.71}$   \\
                                                                   & $0.2,0.8,0.0$   & $\bf{0.30}$   \\
\bottomrule
\end{tabular}
\label{tab:sims2}
\end{table}

A variety of specifications are presented in Tables \ref{tab:sims1} and \ref{tab:sims2}, with specific parameters being varied as others are held constant to demonstrate the impact on the power of the test. For each specification, $5000$ samples are simulated, and for each simulation, a test of the hypothesis that the LATE is zero is undertaken using the Wald IV estimator. This allows for a simulated power to be computed by calculating the proportion of the $5000$ simulations for which it is possible to reject the null hypothesis that $\tau = 0$ given a two-tailed test with a type-I error tolerance of $\alpha = 0.05$. The results shown in Tables \ref{tab:sims1} and \ref{tab:sims2} demonstrate that, in addition to the investigation parameters, all nine of the distribution parameters have an impact on the power of the test.

%\begin{table}[htbp]
%\footnotesize
%\centering
%\caption{LATE Power Simulations (Computed Power in Bold)}
%\begin{tabular}{|p{8cm}|p{2cm}p{2cm}p{2cm}|}
%\hline
%Fixed Parameters: & \multicolumn{3}{|c|}{Varying Parameter} \\
%\hline
%$N = 1000$, $\tau = 5$, $\pi = 0.2$, & \multicolumn{3}{|c|}{Varying $\tau$ = } \\
%$E[Y_C(0)] = 0$, $SD(Y_C(0)) = 8$, $SD(Y_C(1)) = 8$, & $x$ & $x$ & $x$ \\
%$\dot{\pi}_{NT} = 0.4$, $E[Y_{NT}(0)] = -3$, $SD(Y_{NT}(0)) = 12$, & \multicolumn{3}{|c|}{Power ($1-\beta$) = }    \\ 
%$\dot{\pi}_{AT} = 0.4$,  $E[Y_{AT}(1)] = 3$, $SD(Y_{AT}(1)) = 4$& $\bf{0.}$ & $\bf{0.}$ & $\bf{0.}$   \\ \hline
%\end{tabular}
%\label{tab:sims}
%\end{table}

Table \ref{tab:sims3} compares the power of the Wald IV estimator to reject the null hypothesis that the LATE equals zero with the power of the difference-in-means estimator to reject the null hypothesis that the ITT equals zero, highlighting a phenomenon recognized in previous work by \cite{jo2002}. Specifically, Table \ref{tab:sims3} varies the mean outcome values of the never-takers and always-takers, holding all other parameters constant. As can be seen, the power for the two tests changes at different rates as these two principal strata means are altered. Depending upon the specification, the power of the Wald IV estimator may be higher or lower than the power of the ITT difference-in-means estimator. Table \ref{tab:sims3} demonstrates that a power analysis for the ITT cannot serve as a replacement for a LATE power analysis.
%---that is, unless the analyst is willing to make additional and possibly unjustified assumptions about the distribution parameters.

\begin{table}[ht!]
\scriptsize
\centering
\caption{LATE vs. ITT Power Simulations}
\begin{tabular}{p{8cm}ccc}
\toprule
Fixed Parameters: & Varying Parameter(s) & Power$_{LATE}$ & Power$_{ITT}$ \\
\midrule
													  & $E[Y_{NT}(0)], E[Y_{AT}(1)] = $ & \:\:\:\:\:\:\:\:\:\:\:\:\:\:\:\:\:\:\:\:\:\:\:\:\:\:\:\:\:\: & \:\:\:\:\:\:\:\:\:\:\:\:\:\:\:\:\:\:\:\:\:\:\:\:\:\:\:\:\:\: \\
$N = 1000$, $\tau = 5$, $\pi = 0.2$,                               & $-10,10$  & $\bf{0.34}$ & $\bf{0.25}$   \\
$E[Y_C(0)] = 0$, $SD(Y_C(0)) = 8$, $SD(Y_C(1)) = 8$,                  & $-10,3$   & $\bf{0.38}$ & $\bf{0.30}$   \\ 
$\dot{\pi}_{NT} = 0.4$, $SD(Y_{NT}(0)) = 12$,                       & $-3,3$    & $\bf{0.44}$ & $\bf{0.41}$   \\
$\dot{\pi}_{AT} = 0.4$, $SD(Y_{AT}(1)) = 4$                         & $10,3$    & $\bf{0.26}$ & $\bf{0.37}$   \\ 
                                                                   & $10,-6$   & $\bf{0.13}$ & $\bf{0.30}$   \\
\bottomrule
\end{tabular}
\label{tab:sims3}
\end{table}

Finally, Table \ref{tab:sims4} compares the power of the Wald IV estimator to reject the null hypothesis that the LATE equals zero with the power implied by scaling a standard ATE power analysis. As described by previous studies \citep{dufloetal2007,baiocchietal2014}, one approach to performing a power analysis in the face of noncompliance is by performing a simple ATE power analysis but scaling the variance of the standard ATE difference-in-means estimator, specifically dividing it by the compliance rate squared. This implies that, for a given treatment effect, the sample size required to reach a particular power level in the hypothetical full-compliance context must, if there is actually imperfect compliance, then be divided by the compliance rate squared to offer the sample size required to reach that same power level for the LATE given that noncompliance. For instance, if $100$ observations are needed in the ATE context given full compliance, then approximately $\frac{100}{0.2^2} = 2500$ observations are required in the LATE context given a compliance rate of $0.2$. However, this approximate equality is only achieved given the strong assumptions that (a) the never-takers have the same mean outcome value as the untreated compliers, (b) the always-takers have the same mean outcome value as the treated compliers, and (c) all groups have the same within-condition outcome variance. If any of those assumptions are violated, the true power of the Wald IV estimator can diverge dramatically from the power implied by this scaled ATE power analysis. 

Table \ref{tab:sims4} demonstrates this result. Specifically, a standard ATE power analysis is performed given a treatment effect of $5$, a sample size of $100$ (assigned equally to the treatment and control groups), and within-group outcome standard deviation of $8$. The resulting power is $0.87$, as indicated in the last column of the table. The other columns present the results of additional LATE power simulations. For all of the simulations, the treatment effect and within-group variances for the compliers are also set to $5$ and $8$, respectively, while the compliance rate is set to $0.2$. This implies that $\frac{100}{0.2^2} = 2500$ observations are required to reach a power of $0.87$, and thus all simulations are set with a sample size of $2500$. The results in the second and third columns of the table show that the power for the Wald IV estimator is indeed about $0.87$ when the appropriate distributional assumptions are met with regard to the always-takers and never-takers (first specification). However, as the results also show, as soon as those assumptions are violated, the scaled ATE power analysis provides extremely unreliable guidance on power for the LATE.

\begin{table}[ht!]
\scriptsize
\centering
\caption{LATE vs. Scaled ATE Power Simulations}
\begin{tabular}{p{5cm}ccc}
\toprule
Fixed Parameters: & Varying Parameter(s) & Power$_{LATE}$ & Power Implied by \\
                  &                      &                & Scaled ATE Analysis \\
\midrule
                                                                  & $E[Y_{NT}(0)], E[Y_{AT}(1)],$ &   &  \\
                                                                  & $SD(Y_{NT}(0)), SD(Y_{AT}(1)) = $ &   &  \\
$N = 2500$, $\tau = 5$, $\pi = 0.2$,                               & $0,5,8,8$       & $\bf{0.87}$ &  $\cdot$  \\
$E[Y_C(0)] = 0$,                                                    & $-10,15,8,8$    & $\bf{0.54}$ &  $\cdot$  \\ 
$SD(Y_C(0)) = 8$, $SD(Y_C(1)) = 8$,                                  & $0,5,16,16$     & $\bf{0.39}$ & $\bf{0.87}$   \\
$\dot{\pi}_{NT} = 0.4$, $\dot{\pi}_{AT} = 0.4$                     & $-10,15,16,16$  & $\bf{0.32}$ &  $\cdot$ \\ 
                                                                   & $15,-10,16,16$  & $\bf{0.19}$ &  $\cdot$  \\
\bottomrule
\end{tabular}
\label{tab:sims4}
\end{table}

%\begin{table}[ht!]
%\scriptsize
%\centering
%\caption{LATE vs. ITT Power Simulations}
%\begin{tabular}{p{8cm}ccc}
%\toprule
%Fixed parameters: & Varying Parameter(s) & Power$_{LATE}$ & Power$_{ITT}$ \\
%\midrule
%$N = 3000$, $\tau = 5$, $\pi = 0.2$, & $E[Y_{NT}(0)], E[Y_{AT}(1)] = $ & \:\:\:\:\:\:\:\:\:\:\:\:\:\:\:\:\:\:\:\:\:\:\:\:\:\:\:\:\:\: & \:\:\:\:\:\:\:\:\:\:\:\:\:\:\:\:\:\:\:\:\:\:\:\:\:\:\:\:\:\: \\
%$E[Y_C(0)] = 0$, $SD(Y_C(0)) = 8$, $SD(Y_C(1)) = 8$,                  & $-10,10$  & $\bf{0.66}$ & $\bf{0.58}$   \\
%$\dot{\pi}_{NT} = 0.4$, $SD(Y_{NT}(0)) = 12$,                       & $-10,3$   & $\bf{0.76}$ & $\bf{0.71}$   \\ 
%$\dot{\pi}_{AT} = 0.4$, $SD(Y_{AT}(1)) = 4$                         & $-3,3$    & $\bf{0.86}$ & $\bf{0.84}$   \\
%                                                                   & $10,3$    & $\bf{0.79}$ & $\bf{0.83}$   \\ 
%                                                                   & $10,-6$   & $\bf{0.59}$ & $\bf{0.67}$   \\
%\bottomrule
%\end{tabular}
%\label{tab:sims3b}
%\end{table}

\clearpage

\section*{Appendix C: Examples of Ordered Means Assumption ($\bar{Y}_{NT} \leq \bar{Y}_{C} \leq \bar{Y}_{AT}$)}

\setcounter{table}{0}
\renewcommand{\thetable}{C\arabic{table}}%

\vspace{1cm}

\subsection*{Example 1: National JTPA Study \citep{bloometal1997}}

\begin{table}[ht!]
\footnotesize
\centering
\caption{Example 1 Results by Treatment Assignment and Uptake}
\begin{tabular}{c|cc}
\multicolumn{1}{c}{} & $\mathbf{D_i=0}$ & $\mathbf{D_i=1}$ \\
      \cline{2-3} \\
$\mathbf{Z_i=0}$ & $N=3663$              & $N=54$     \\
               & $\bar{Y} = 15062.99$  & $\bar{Y} = 13515.26$     \\ \\
$\mathbf{Z_i=1}$ & $N=2683$              & $N=4804$    \\
               & $\bar{Y} = 13979.93$  & $\bar{Y} = 17439.8 $     \\ \\

\end{tabular}
\label{tab:jtpa}
\end{table}

This example pertains to a well-known experimental study of the Job Training Partnership Act of 1982 \citep{bloometal1997}, where study subjects were randomly assigned such that they were either allowed to enroll in a job training program (assigned to the treatment) or excluded from the training for an 18-month period (assigned to control). The dataset used here is the tabulation of the JTPA study data by \cite{abadieetal2002}. The data correspond to adult participants in the JTPA experiment for whom there was a measurement of the outcome variable, which was earnings in the 30-month period after random assignment. Table \ref{tab:jtpa} separates the sample into four groups based on treatment assignment ($Z$) and uptake ($D$). As shown in Table \ref{tab:jtpa}, there were almost no identified always-takers ($Z_i=0,D_i=1$) in the National JTPA Study. As explained in Lemma \ref{narrow}, the alternative power lower bound can be used when $Cov(E[D_i|X_i],E[Y_i|X_i])$ is non-negative, where $X_i$ denotes the principal strata, which is true given the ordered means assumption that $\bar{Y}_{NT} \leq \bar{Y}_{C} \leq \bar{Y}_{AT}$. However, given how few always-takers there are, they will have a negligible influence on $Cov(E[D_i|X_i],E[Y_i|X_i])$, and hence the ordered means assumption can be simplified to $\bar{Y}_{NT} \leq \bar{Y}_{C}$ as a practical matter.

There are 2683 identified never-takers ($Z_i=1,D_i=0$) in the study. Furthermore, this group of identified never-takers exhibits the lowest mean observed outcome among the three groups not including the identified always-takers. As a result, it is reasonable to conclude that the ordered means assumption was met in this study.

\clearpage

\subsection*{Example 2: New Haven Vote Canvassing Experiment \citep{gerbergreen2000,gerbergreen2005}}

\begin{table}[ht!]
\footnotesize
\centering
\caption{Example 2 Results by Treatment Assignment and Uptake}
\begin{tabular}{c|cc}
\multicolumn{1}{c}{} & $\mathbf{D_i=0}$ & $\mathbf{D_i=1}$ \\
      \cline{2-3} \\
$\mathbf{Z_i=0}$ & $N=10733$              & $N=0$     \\
               & $\bar{Y} = 0.442$  & $\bar{Y} = NA$     \\ \\
$\mathbf{Z_i=1}$ & $N=1781$              & $N=869$    \\
               & $\bar{Y} = 0.435$  & $\bar{Y} = 0.577$     \\ \\

\end{tabular}
\label{tab:newhaven}
\end{table}

This example pertains to a voter mobilization field experiment conducted by \cite{gerbergreen2000} in 1998, using updated data from \cite{gerbergreen2005}. The experiment involved three different types of voter mobilization treatments---mailings, phone calls, and home visits. The results presented here include those subjects assigned to receive no treatments (pure control) and those subjects assigned only to the home visits, as this was the treatment in which a major factor was noncompliance---i.e. not actually being contacted as a result of the subject not being home. Subjects for whom the outcome (having voted in the 1998 midterm election) was missing are omitted.

As Table \ref{tab:newhaven} shows, there is strict one-sided noncompliance by design in this study, with no always-takers, thus simplifying the ordered means assumption to $\bar{Y}_{NT} \leq \bar{Y}_{C}$. The results are consistent with this assumption, as the identified never-takers ($Z_i=1,D_i=0$) have the lowest mean outcome value.

\clearpage

\subsection*{Example 3: Fuzzy Regression Discontinuity Design on the Effect of Naturalization on Integration \citep{hainhanpiet2015}}

\begin{table}[ht!]
\footnotesize
\centering
\caption{Example 3 Results by Treatment Assignment and Uptake}
\begin{tabular}{c|cc}
\multicolumn{1}{c}{} & $\mathbf{D_i=0}$ & $\mathbf{D_i=1}$ \\
      \cline{2-3} \\
$\mathbf{Z_i=0}$ & $N=69$              & $N=90$     \\
               & $\bar{Y} = -0.527$  & $\bar{Y} = 0.048$     \\ \\
$\mathbf{Z_i=1}$ & $N=0$              & $N=244$    \\
               & $\bar{Y} = NA$  & $\bar{Y} = 0.055$     \\ \\

\end{tabular}
\label{tab:naturalization}
\end{table}

This example pertains to a study on the effects of naturalization on the political integration of immigrants who applied for Swiss citizenship prior to 2003 \citep{hainhanpiet2015},\footnote{The data are not publicly available due to privacy concerns, and the statistics in Table \ref{tab:naturalization} were provided by the authors of the study. Individuals interested in accessing the data for replication purposes should contact the authors.} during which time certain municipalities used secret ballot referendums to vote on which immigrants should be granted citizenship. As one of the identification and estimation strategies, the study includes a fuzzy regression discontinuity design comparing only those immigrants who barely received a majority ``yes" vote and those who barely missed the majority. However, some of those immigrants who did not receive the majority vote initially went on to gain citizenship later. As a result, a fuzzy regression discontinuity design can be applied, where $Z$ indicates having received the ``yes" vote (assignment to treatment) and $D$ indicates having gained citizenship (uptake of the treatment).

The fuzzy regression discontinuity design (RDD) combines the machinery of the sharp RDD and the LATE. The method of power analysis presented in this study would be applicable to one particular estimator sometimes used with the fuzzy RDD, which is the standard Wald IV estimator applied to subjects within the specified RDD bandwidth. However, while the power method is technically not applicable to the actual fuzzy RDD regression estimators, it could be used as a convenient proxy.

As Table \ref{tab:naturalization} shows, there is strict one-sided noncompliance by design in this study, with no never-takers, thus simplifying the ordered means assumption to $\bar{Y}_{C} \leq \bar{Y}_{AT}$. The identified always-takers ($Z_i=0,D_i=1$) have almost the highest mean outcome value. However, the mean outcome of the compliers would be the weighted average of the mean outcome of the identified compliers in the ($Z_i=0,D_i=0$) group and the unidentified compliers in the ($Z_i=1,D_i=1$). While we cannot identify the specific units in the latter group that are compliers, rather than always-takers, we have information on the identified always-takers ($Z_i=0,D_i=1$) and knowledge that the ($Z_i=1,D_i=1$) group is a mix of compliers and always-takers, which allows us to assess the ordered means assumption.

%Based on that information, the mean outcome value of the unidentified compliers in the ($Z_i=1,D_i=1$) group is unlikely to be much higher, if at all, than the mean outcome of the identified always-takers ($Z_i=0,D_i=1$). 
Specifically, treating $Z$ as if randomized, we would expect the proportion of individuals who are always-takers to be equal across the ($Z_i=0$) and ($Z_i=1$) margins. Thus, we can estimate that approximately 138 of the individuals in the ($Z_i=1,D_i=1$) group are always-takers, whose mean outcome should be the same (in expectation) as the mean outcome of the identified always-takers in the ($Z_i=0,D_i=1$) group, which is 0.048. This implies that the mean outcome of the unidentified compliers in the ($Z_i=1,D_i=1$) group, of which we expect there should be 106, is approximately 0.064. As a result, the mean outcome of all compliers would undoubtedly be lower than the mean outcome of always-takers, given the particularly low value of the mean outcome for the identified compliers in the ($Z_i=0,D_i=0$) group. Specifically, using the information we have, we can estimate the mean outcome of all compliers to be $-0.17$. Thus, the results are consistent with the ordered means assumption.

\clearpage

\section*{Appendix D: Power Plots}

\setcounter{figure}{0}
\renewcommand{\thefigure}{D\arabic{figure}}%

\begin{figure}[ht!]
\begin{center}
\caption{Power by Effect Size (with Varying Compliance Rate)} \label{fig:powerbykappa}
\includegraphics[scale=0.35]{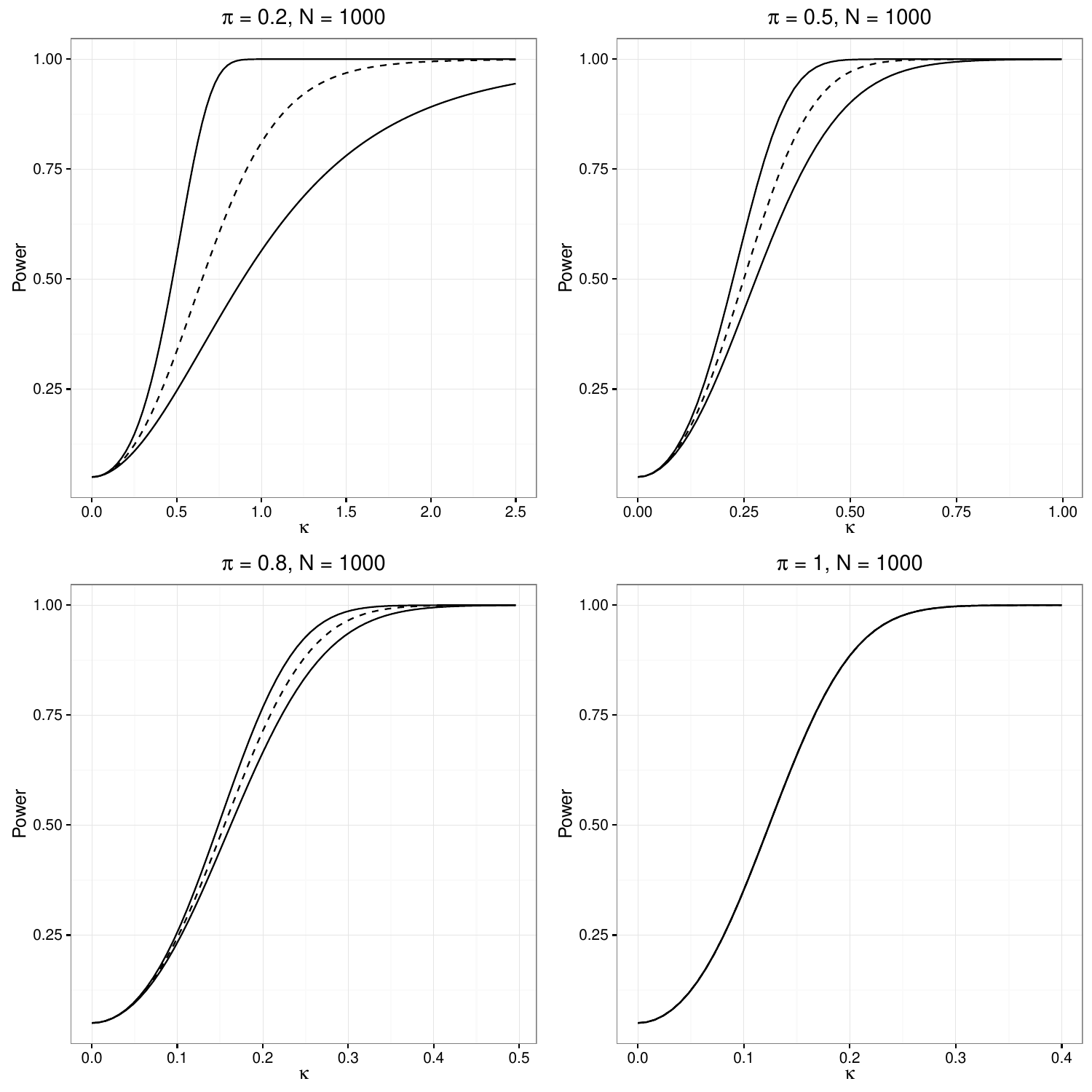}
\end{center}
\end{figure}

\begin{figure}[ht!]
\begin{center}
\caption{Power by Sample Size (with Varying Compliance Rate)} \label{fig:powerbyn}
\includegraphics[scale=0.35]{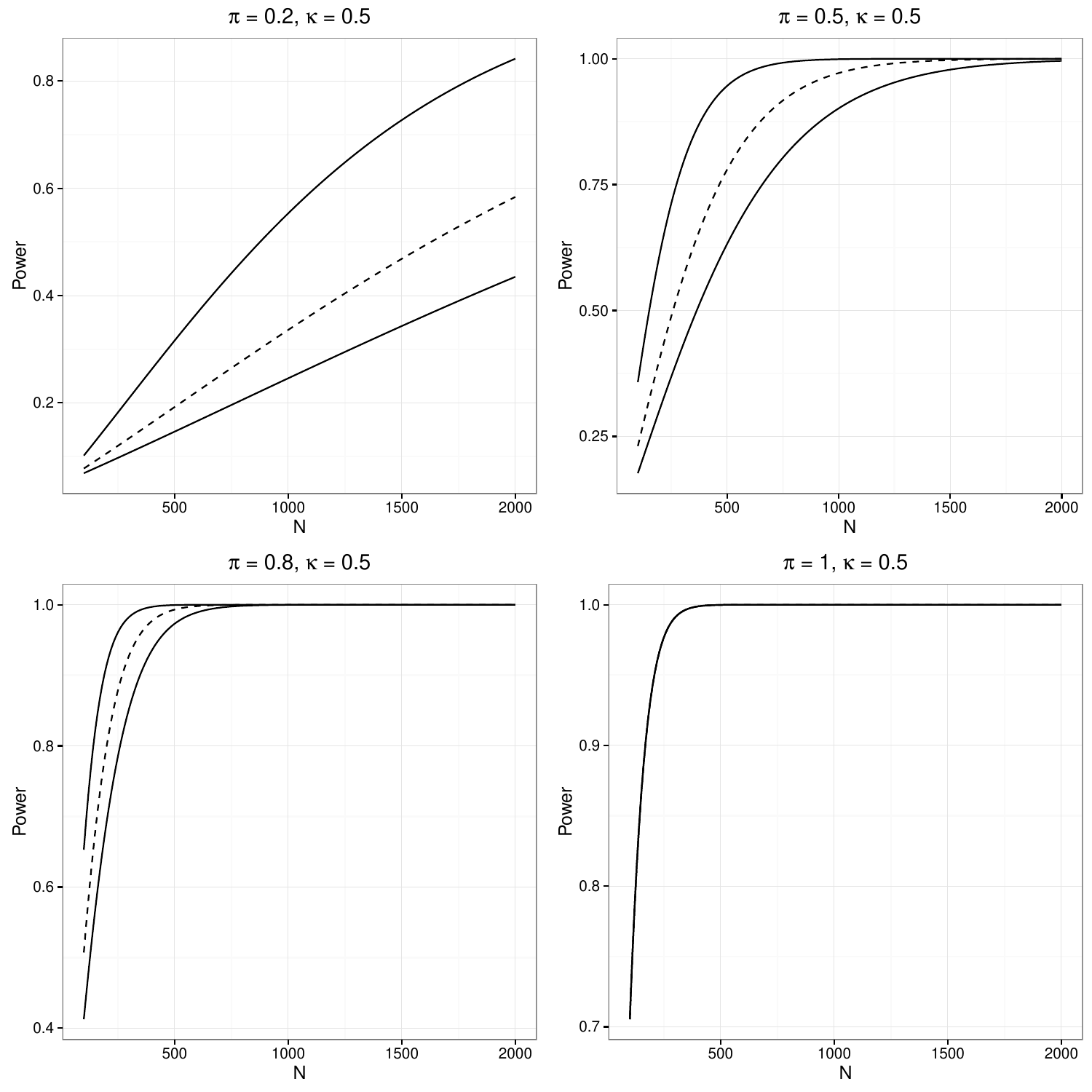}
\end{center}
\end{figure}

\begin{figure}[ht!]
\begin{center}
\caption{MDES by Sample Size (with Varying Compliance Rate)} \label{fig:kappabyn}
\includegraphics[scale=0.35]{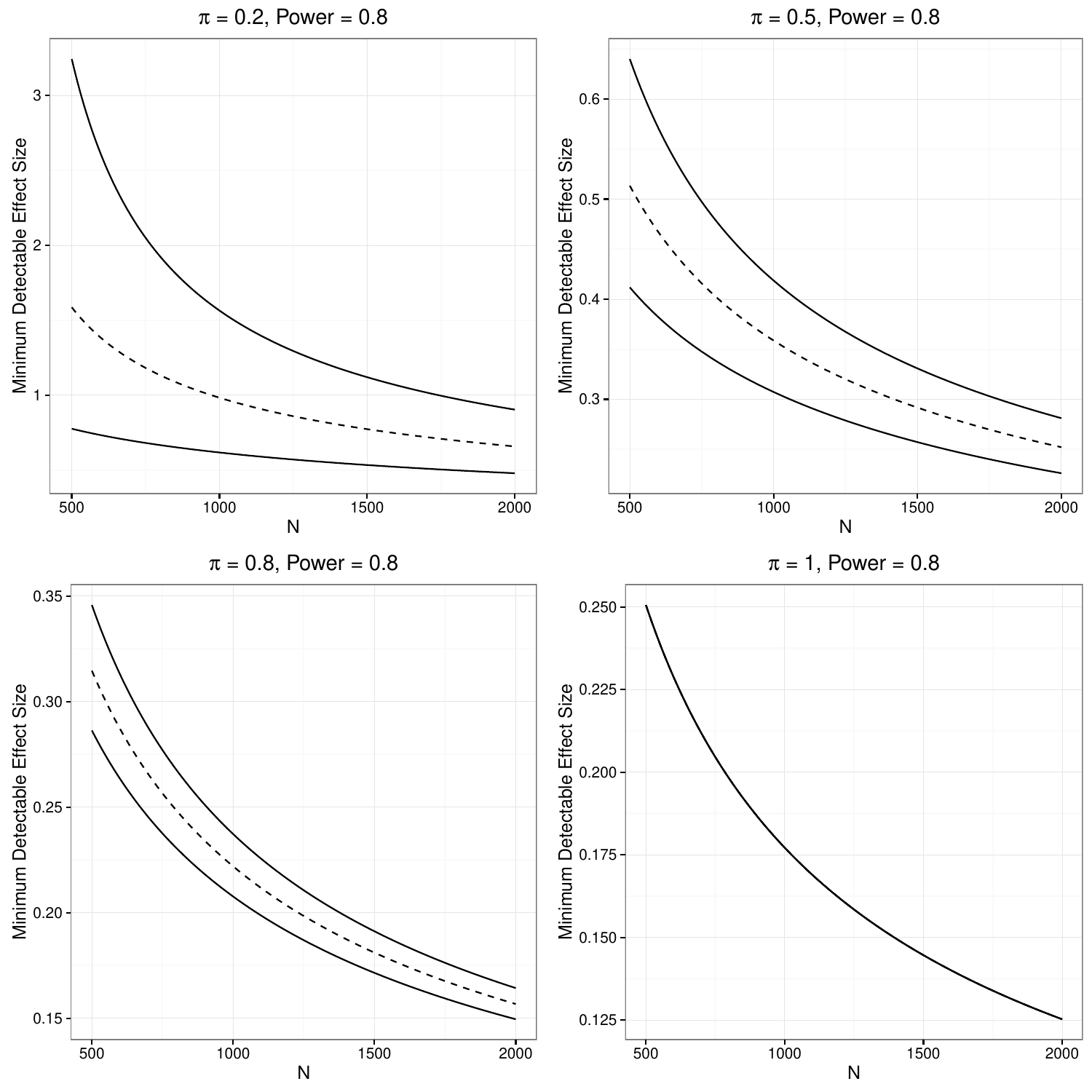}
\end{center}
\end{figure}

\begin{figure}[ht!]
\begin{center}
\caption{Sample Size by Effect Size (with Varying Compliance Rate)} \label{fig:nbykappa}
\includegraphics[scale=0.35]{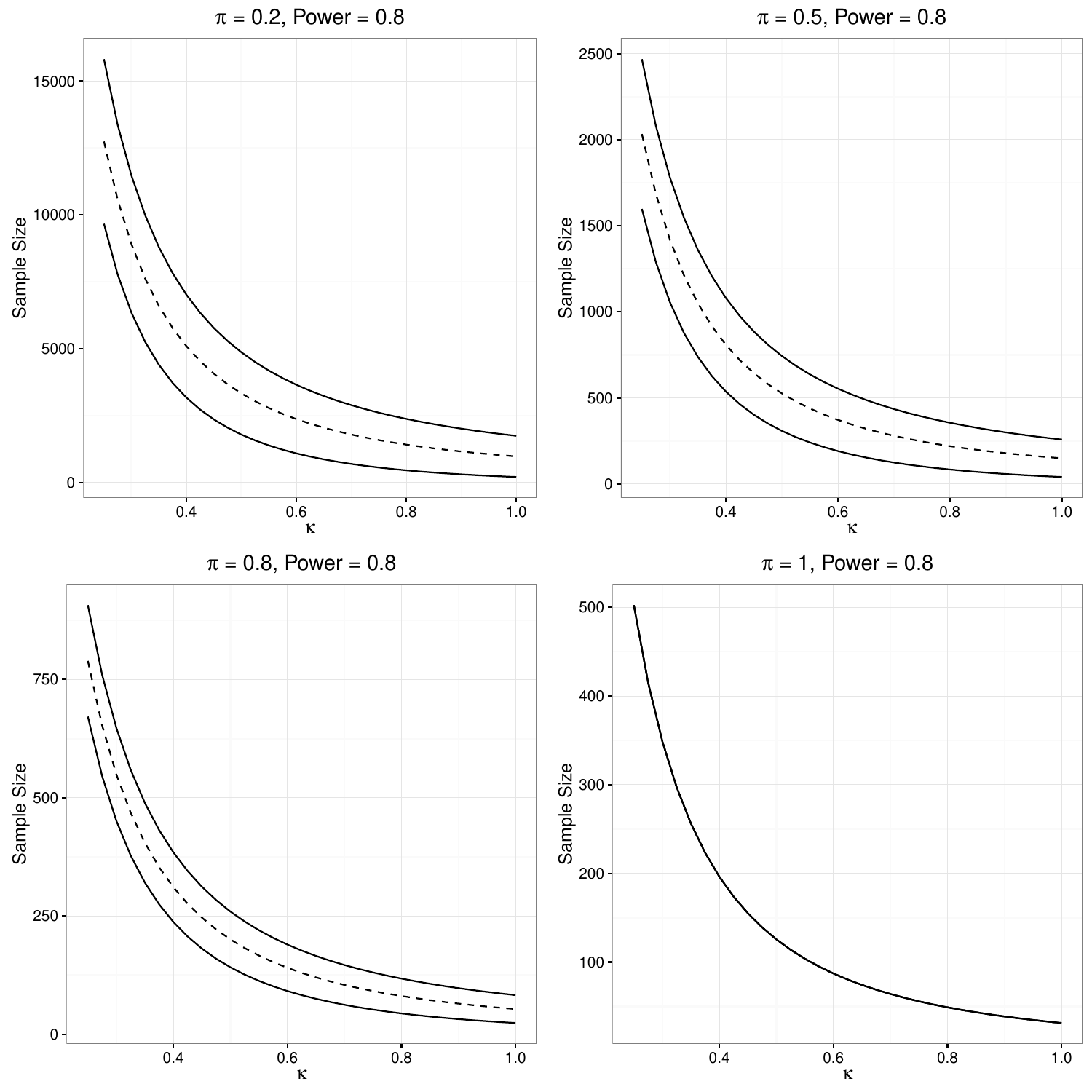}
\end{center}
\end{figure}

This section provides graphical illustrations of the relationships between power and the investigation parameters in the LATE power analysis method presented in this study. Figures \ref{fig:powerbykappa}-\ref{fig:nbykappa} each contain a set of four plots. For each set, the $x$-axis plots a particular investigation parameter while the $y$-axis plots either power or another investigation parameter. In addition, the compliance rate is varied across the four plots in each set. Every plot contains three curves. The two solid curves correspond to the lower and upper bounds. The dashed line corresponds to the alternative conservative bound under Assumption \ref{assump:om} (ordered means). The type-I error tolerance ($\alpha$) is held constant at $0.05$ for all plots. 

Figure \ref{fig:powerbykappa} plots the power against the effect size ($\kappa$). The sample size ($N$) is set to 1000 for all four plots in this figure, while the compliance rate ($\pi$) varies between 0.2, 0.5, 0.8, and 1. In addition to showing the general rate at which power increases as the effect size increases, the figure illustrates the importance of the compliance rate in shaping the relationship between power and the effect size. First, it shows that the rate at which power increases with the effect size is faster for higher compliance rates. For instance, for the alternative lower bound (the dashed line), given $\pi=0.2$, a power of 0.8 is achieved only when $\kappa \approx 1$. In contrast, this power is achieved when $\kappa \approx 0.25$ given $\pi = 0.8$. In addition, the figure also shows that the bounds converge as $\pi \rightarrow 1$. When $\pi = 1$ (i.e. full compliance), the LATE becomes the ATE, which does not require bounds to characterize the power.

Figure \ref{fig:powerbyn} plots the power against the sample size ($N$), with $\kappa$ held constant and $\pi$ varying across plots as previously. As in Figure \ref{fig:powerbykappa}, Figure \ref{fig:powerbyn} shows a similar impact of the compliance rate on the relationship between the sample size and power. Figures \ref{fig:kappabyn} and \ref{fig:nbykappa} plot the relationship between effect size and sample size, given a prespecified desired level of power (held constant at 0.8). Again, the compliance rate varies across plots, illustrating the importance of the compliance rate in determining the relationship between the other parameters and the distance between the bounds.

\clearpage

\section*{Appendix E: Simulation Results Given $P(Z_i=1) = 0.25$}

\setcounter{figure}{0}
\renewcommand{\thefigure}{E\arabic{figure}}%

\begin{figure}[ht!]
\begin{center}
\caption{Simulated Power vs. Analytic Bounds, $P(Z_i=1)=0.25$} \label{fig:simsboundsZrelaxed}
\includegraphics[scale=0.55]{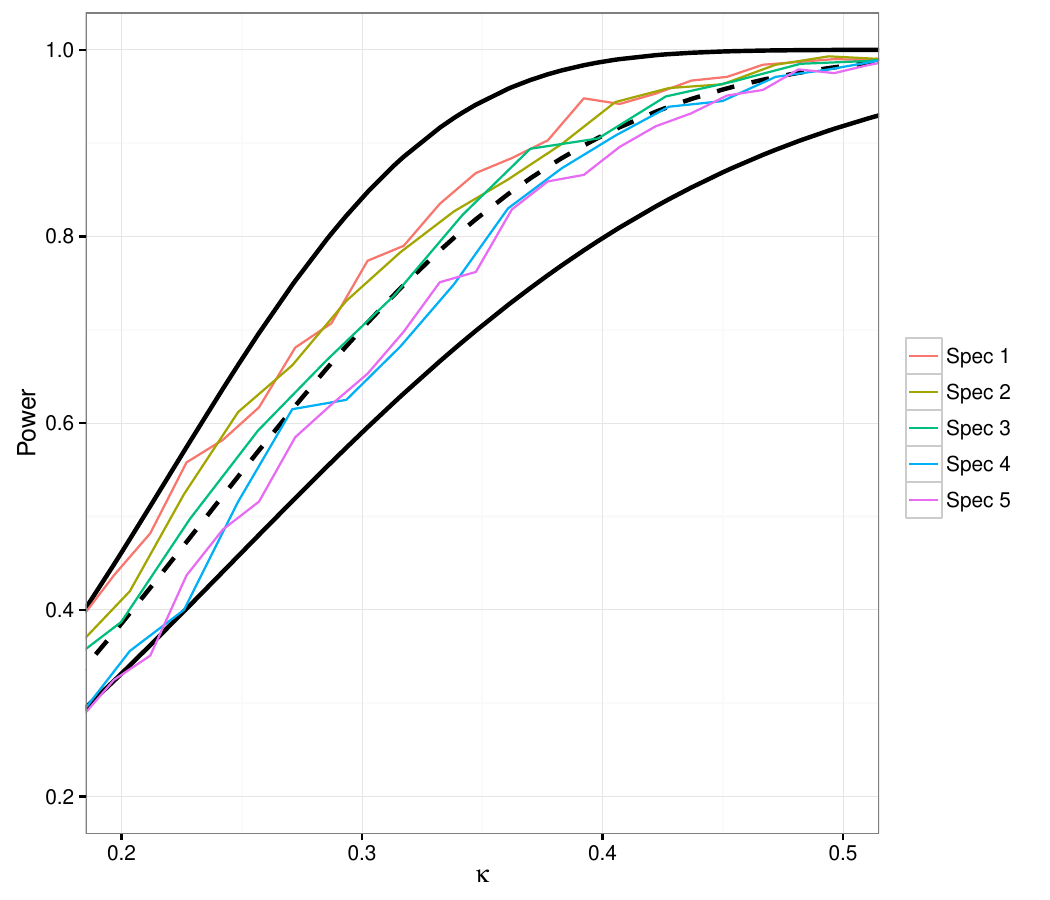}
\end{center}
\end{figure}

Figure \ref{fig:simsboundsZrelaxed} compares simulated power curves to the analytic bounds given $P(Z_i=1) = 0.25$, where power is plotted as a function of $\kappa$. As in the simulations presented in the main text, the simulations presented here also each specify a data-generating distribution of the quadruples $(Y_i(0),Y_i(1),D_i(0),D_i(1)) \in \mathbb{R} \times \mathbb{R} \times \{0,1\} \times \{0,1\}$, randomly draw from that distribution, and randomize the treatment assignment variable, generating samples of independent and identically distributed units of the form $(Y_i, D_i, Z_i) \in \mathbb{R} \times \{0,1\} \times \{0,1\}$. As previously, the solid black lines denote the analytic upper and lower bounds, while the dashed black line denotes the alternative lower bound under the ordered means assumption. The colored lines denote power curves that were simulated by specifying the full set of investigation and distribution parameters, and hence simulating all three principal strata. The specifications are the same as those pertaining to Figure \ref{fig:simsbounds} in the main text; the only difference is that $P(Z_i=1) = 0.25$, rather than $P(Z_i=1) = 0.5$.

As Figure \ref{fig:simsboundsZrelaxed} illustrates, the analytic bounds derived for the general case where $P(Z_i=1) = p_z$ perform similar to the bounds derived for the special case of $P(Z_i=1) = 0.5$.
%\footnote{Again, the simulated power curves fall slightly outside the bounds at low levels of power. As described earlier, this is not necessarily surprising given that the power formula relies upon asymptotic normality, the LATE estimator is likely to exhibit somewhat irregular tail behavior in finite samples, and the LATE estimator's variance must itself be estimated.
%whereas the simulations are based upon finite samples and hence deal with random variables not exactly distributed standard normal (the tails are probably fatter).
%} 
Compared to Figure \ref{fig:simsbounds}, the power curves (analytic and simulated) in Figure \ref{fig:simsboundsZrelaxed} are all slightly lower, which makes sense given that the decrease in $P(Z_i=1)$ means a less balanced allocation of treatment uptake among the compliers. Similar to Figure \ref{fig:simsbounds}, the alternative lower bound (the dashed black line) in Figure \ref{fig:simsboundsZrelaxed} bounds the appropriate simulated curves. For specifications 1 and 2, Assumption \ref{assump:om} (ordered means) is met at all values of $\kappa$, and hence the alternative lower bound applies. Accordingly, the curves for these specifications again lie above the alternative lower bound. In contrast, the ordered means assumption is violated by specifications 4 and 5 at all values of $\kappa$, and again these curves appropriately lie below the alternative lower bound.

\clearpage

\section*{Appendix F: Incorporating Covariates}

The bounds on the variance of the estimator presented in Lemma \ref{step3} (see Appendix A) provide a convenient framework for incorporating the gains in precision achieved through covariate adjustment, which can be quantified in terms of $R^2$ values, as presented elsewhere \citep[e.g.][]{bloometal2007}.

As presented in the main body of the study, define the following $R^2$ values:
$$R^2_{DW} = \frac{\sigma^2 - \sigma^{*2}}{\sigma^2}$$
$$R^2_{YW} = \frac{\omega^2 - \omega^{*2}}{\omega^2}$$
where $\sigma^{2} = E[\nu_i^{2}]$ as defined in the proof of Proposition \ref{step6}, $\omega^{2} = E[\zeta_i^{2}]$ as defined in the proof of Proposition \ref{step6}, $\sigma^{*2} = E[\nu_i^{*2}]$ from equation (\ref{eq1cov}), and $\omega^{*2} = E[\zeta_i^{*2}]$ from equation (\ref{eq3cov}).

$R^2_{DW}$ measures the proportion of variation in $D$ left unexplained by $Z$ that is explained by the covariates contained in $\mathbf{W}$, while $R^2_{YW}$ measures the proportion of variation in $Y$ left unexplained by $Z$ that is explained by the covariates contained in $\mathbf{W}$. This definition can also be re-expressed as $\sigma^{*2} = (1 - R^2_{DW}) \sigma^{2}$ and $\omega^{*2} = (1 - R^2_{YW}) \omega^{2}$, where $\sigma^{*2}$ and $\omega^{*2}$ measure the remaining unexplained variation in $D$ and $Y$ after adjusting for $Z$ and $\mathbf{W}$.

The result is that, if a researcher plans to measure relevant covariates and has knowledge about anticipated $R^2_{DW}$ and $R^2_{YW}$ values based on previous research or subject matter expertise, these values can then be used to further modify the power bounds of the 2SLS estimator of the LATE. Specifically, the variance bounds in Lemma \ref{step3} would now be based on $\sigma^{*2}$ and $\omega^{*2}$ rather than $\sigma^{2}$ and $\omega^{2}$. That is, the bounds in Lemma \ref{step3} become:
$$V_N^{\widehat{2SLS}} \: bounds: \:\:\:\: \frac{\omega^{*2} + \tau^2 \sigma^{*2} \pm 2 \tau \sigma^{*} \omega^{*}}{0.25 N \pi^2}$$  
$$= \frac{(1-R^2_{YW})\omega^{2} + \tau^2 (1-R^2_{DW}) \sigma^{2} \pm 2 \tau \sqrt{(1-R^2_{DW})(1-R^2_{YW})} \sigma \omega}{0.25 N \pi^2}$$
$$= \frac{(1-R^2_{YW})E[\zeta_i^2] + \tau^2 (1-R^2_{DW}) E[\nu_i^2] \pm 2 \tau \sqrt{(1-R^2_{DW})(1-R^2_{YW})E[\zeta_i^2]E[\nu_i^2]}}{0.25 N \pi^2}$$
In contrast, $\omega^2 = E[\zeta_i^2]$ is not replaced by $\omega^{*2} = E[\zeta_i^{*2}]$ in Lemma \ref{step5} because the results of Lemma \ref{step5} continue to follow from conditioning on $Z$ alone. The resulting power bounds can then be derived as before, now simply taking into account the scalars $(1 - R^2_{DW})$ and $(1 - R^2_{YW})$ introduced into the bounds on the estimator variance. Continuing to assume without loss of generality that $\kappa > 0$ (and $\tau > 0$) is under investigation, the result is the following:

$$\sqrt{ \frac{0.25 \kappa^2 \pi^2 N}{(1-R^2_{YW}) + \kappa^2 (1-R^2_{DW}) E[\nu_i^2] + 2 \kappa \sqrt{(1-R^2_{YW})(1-R^2_{DW}) E[\nu_i^2]} } }$$
$$\leq \frac{\tau}{\sqrt{V_N^{\widehat{2SLS}}}}$$
$$\leq \sqrt{ \frac{0.25 \kappa^2 \pi^2 N}{(1-R^2_{YW}) + \kappa^2 (1-R^2_{DW}) E[\nu_i^2] - 2 \kappa \sqrt{(1-R^2_{YW})(1-R^2_{DW}) E[\nu_i^2]} } }$$

%$$\frac{\tau}{\sqrt{V_N^{\widehat{2SLS}}}} \: bounds:  \:\:\:\: \sqrt{ \frac{0.25 \kappa^2 \pi^2 N}{(1-R^2_{YW}) + \kappa^2 (1-R^2_{DW}) E[\nu_i^2] \pm 2 \kappa \sqrt{(1-R^2_{YW})(1-R^2_{DW}) E[\nu_i^2]} } }$$

The bounds can be plugged into the power formula $\Phi \left(-c^* + \frac{\tau}{\sqrt{V_N^{\widehat{2SLS}}}}\right) + \Phi \left(-c^* - \frac{\tau}{\sqrt{V_N^{\widehat{2SLS}}}}\right)$. As previously, the bounds can be modified to both relax Assumption \ref{assump:epa} (equal assignment probability) and employ Assumption \ref{assump:om} (ordered means), and $E[\nu_i^2]$ replaced with either $(0.5 - \frac{\pi}{2})(0.5 + \frac{\pi}{2})$ or $0.25$ depending on whether Assumption \ref{assump:epa} is made.

\clearpage

\addtolength{\baselineskip}{-0.1\baselineskip}
\bibliographystyle{apalike}
\bibliography{references}

\end{document}